\documentclass[aps, reprint, nofootinbib, longbibliography, superscriptaddress, floatfix]{revtex4-1}

\usepackage{amsmath, amssymb, amsthm}
\usepackage{bm,bbm}

\usepackage{hyperref}
\usepackage[utf8]{inputenc}

\usepackage{soul}
\usepackage{url}
\usepackage{mathtools}
\usepackage{braket}
\usepackage{dsfont}
\usepackage[english]{babel}
\usepackage{graphicx} 
\usepackage{tikz}
\usepackage{float}
\usetikzlibrary{quantikz}

\usepackage{hyperref}
\usepackage{natbib} 
\usepackage{physics}

\usepackage{listings}
\usepackage{xcolor}
\usepackage{amssymb}

\setbox0=\vbox to \vsize{\unvbox255\vfill}

\usepackage{multirow}
\usepackage{tabu}
\usepackage{makecell}
\usepackage{color}

\usepackage{lipsum,parskip,booktabs,graphicx,tabulary,tabularx}
\usepackage{tabu}

\usepackage{wrapfig}
\newtheorem{definition}{Definition}

\newcommand{\LDIGUS}{
Lighthouse Disruptive Innovation Group, LLC,
7 Broadway Terrace, Apt 1,
Cambridge MA 02139,
Middlesex County, Massachusetts (USA)
}

\newcommand{\LDIGEU}{
Lighthouse Disruptive Innovation Group Europe, SL.
Barcelona - Spain
}

\newcommand{\DSD}{
Smart Society Research Group -
La Salle - Universitat Ramon Llull,
Carrer de Sant Joan de La Salle, 42,
08022 Barcelona (Spain)
}

\newcommand{\MIT}{
MIT Media Lab - City Science Group, Cambridge, USA
}

\usepackage{babel}

\graphicspath{{./images/}}

\begin{document}
\title{Quantum improvement in Spatial Discretization}

\author{Saul Gonzalez}
\affiliation{\LDIGEU}
\email{saul.gonzalez@lighthouse-dig.com}

\author{Parfait Atchade-Adelomou}
\affiliation{\MIT}
\email{parfait@mit.edu}
\affiliation{\DSD}
\affiliation{\LDIGUS}

\email{parfait.atchade@lighthouse-dig.com}


\date{November 2023}

\begin{abstract}
Quantum algorithms have begun to surpass classical ones in several computation fields, yet practical application remains challenging due to hardware and software limitations. Here, we introduce a quantum algorithm that quadratically improves spatial discretization within these constraints. Implemented in the quantum software library Pennylane, our algorithm bridges the gap from theoretical models to tangible quantum circuitry. The approach promises enhanced efficiency in quantum spatial analysis, with simulations and hardware experiments validating its potential.


\end{abstract}
\maketitle

\section{Introduction}\label{sec:intro}

The ascent of quantum computing marks a pivotal turn in computational sciences, heralding a new era primed to redefine the paradigms of data processing and analytical methodologies. Central to this scientific renaissance is the algorithm we introduce crafted to master the discretization of continuous spaces, a ubiquitous challenge intersecting numerous scientific domains \cite{skeel1990method,lustgarten2008improving,grover2021adaptive}.

Transcending classical paradigms, our algorithm operates within the pragmatic bounds of current quantum hardware infrastructures \cite{PhDParfait}. We trace the algorithm's journey from theoretical inception to empirical execution, transmuting its complex underpinnings into quantum simulations. This seminal contribution is a conduit between the conceptual and the practical, fostering real-world quantum applications.

Our focus bifurcates into two distinct yet interlinked objectives: the imperative task of spatial discretization \cite{skeel1990method}, which underpins machine learning and spatial trajectory analysis, and the random and uniform discretization of the feasible region \( \Omega_f \), articulated as \( \Omega_f := R \setminus (\bigcup_{i=1}^m \omega_i) \), with \( \{\omega_i\}_{i=1}^m \) as a set of areas in this space. In veering from classical approaches reliant on discriminator function \( f \) for point inclusion within \( \Omega_f \), we unveil a path less trodden that enhances efficiency and accuracy.

This treatise not only delineates the quantum computational strides enabled by our algorithm but also amplifies its far-reaching implications for real-world scenarios \cite{osher2003geometric}. From the geometric precision required in urban development and vehicular autonomy to the intricate simulations integral to medical and machine learning breakthroughs, our algorithm is set to significantly refine the process of transmuting the continuous into the discrete, outstripping its classical antecedent \cite{sanyal2010almost,funke2016collision, molina2016influence, cao2014spatial,fiorini2022discretization}. With the rise of quantum computing \cite{PhDParfait, a14070194,ParfaitVQE, gonzalez2022gps, atchadeadelomou2023fourier, parfait2020using,consulpacareu2023quantum,atchade2022quantum, alonsolinaje2021eva,adelomou2020formulation,adelomou2023quantum,adelomou2020using}, the prospect of more precise, expedient, and efficient spatial discretization beckons, propelling an array of vital disciplines into an epoch rich with computational potential and innovation.

In this work, we delve into the realm of quantum spatial discretization, aiming to unravel its complexities, enhance its methodologies, and tailor its implementation for the burgeoning \textit{Intermediate-Scale Quantum} (ISQ) era \cite{FromNISQtoISQs}. Our expedition is dual-pronged: scrutinizing and refining spatial discretization's quantum operations and proffering novel implementations suited for the \textit{ISQ} milieu. Through this exploration, we endeavor to strengthen the nexus between quantum theoretical frameworks and their palpable applications, equipping both fledgling and seasoned quantum researchers with cutting-edge tools for spatial discretization.

The manuscript methodically presents research on quantum computing's role in spatial discretization. It begins by outlining the motivations and challenges in quantum computational geometry (Section~\ref{sec:motivation}). It then reviews existing works to contextualize the study's contributions (Section~\ref{sec:workcontext}). Before defining our methodology, we define the problem to solve in Section \eqref{sec:problem}, followed by a foundation of essential quantum concepts applied to geometric problems (Section~\ref{sec:Preliminaries}).  The method is detailed in Section~\ref{sec:IMPLEMENT}, describing the development of quantum circuits, operators, and oracles. Section~\ref{sec:results} validates the theoretical models with empirical evidence, while the concluding Section~\ref{sec:conclusions} reflects on the findings and suggests future research directions, marking the paper's progression from quantum theories to their application in computational geometry.

\section{Motivation and Problem Statement}\label{sec:motivation}

The advent of quantum computing heralds a transformative era in computational methodologies, extending the frontier beyond what classical algorithms can conceive. This research is galvanized by the potential to leverage quantum mechanics to unravel and navigate the complex terrain of geometric problem-solving within the extant landscape of quantum technology.

The enigma of spatial discretization looms large in this new paradigm. Whereas conventional algorithms offer a sturdy foundation, they falter in capturing geometric spaces' intricacy and subtlety, a gap that quantum computing is poised to fill \cite{grover2021adaptive}. Our ambition is to refine and extend classical techniques \cite{braun1997modelling,kummer2017extended}, offering innovative solutions for discretizing spaces interspersed with rectilinear, radial, or polygonal obstacles.

We posit that an algorithmic framework rooted in quantum principles will catalyze a paradigm shift, fostering a sophisticated and adroit analysis of the feasible region \( \Omega_f \). 
To this end, we have implemented and rigorously tested our algorithm within \textit{PennyLane} \cite{bergholm2022pennylane}. 
This discourse is not merely about the theoretical acumen of our algorithm but its ascendancy into the efficiency of quantum computational geometry, heralding a new epoch of algorithmic prowess.

\section{Work Context}\label{sec:workcontext}

Spatial discretization, a cornerstone of computational analysis, has historically been navigated through classical algorithms \cite{braun1997modelling,kummer2017extended}. These algorithms have provided robust frameworks for representing geometric spaces in various applications, including urban planning, computer graphics, and autonomous vehicle navigation. However, they often encounter challenges in accurately capturing the complexity and nuances of geometric configurations. This is where quantum computing's potential becomes particularly compelling \cite{grover2021adaptive}. Our research aims to harness quantum advancements to refine and extend these classical techniques, innovating in spatial discretization amidst diverse obstacles—rectilinear, radial, or polygonal.

In the evolving landscape of quantum computing, significant research has been dedicated to enhancing spatial search algorithms. A notable contribution from Ref. \cite{aaronson2005quantum} delves into the application of Grover's quantum search algorithm \cite{grover1996fast} in spatial contexts, mainly focusing on grid structures. The authors challenge previous assertions, demonstrating that it is feasible to search a \( d \)-dimensional hypercube in \( O(\sqrt{n}) \) time for dimensions \( d \geq 3 \), and in \( O(\sqrt{n} \log^{5/2} n) \) for a two-dimensional space. This work introduces a novel model of quantum query complexity on graphs, incorporating considerations from black hole thermodynamics and the holographic principle, and provides upper and lower bounds for various search tasks. Additionally, it proposes an \( O(\sqrt{n}) \)-qubit communication protocol for the disjointness problem, enhancing previous upper bounds and aligning with quantum communication complexity lower bounds.

Another groundbreaking study on quantum spatial search on a random temporal network \cite{Chakraborty_2017}, which investigates the performance of the \textit{continuous-time quantum walk} (CTQW) algorithm in locating a marked node on a random temporal network composed of Erdős-Rényi random graphs \( G(n; p) \) \cite{renyi1959random}. This research analytically demonstrates that for any given probability \( p \), there exists an interval of time \( \tau \) values for which the algorithm's running time is optimal \( O(\sqrt{n}) \), irrespective of the temporality of the network. The study reveals a threshold \( p_{\text{temp}} = \frac{\log(n)}{\sqrt{n}} \) where the algorithm is optimal regardless of \( \tau \), showcasing that high temporality can lead to optimal performance even when \( p \) is below the static percolation threshold. The paper also discusses applying these findings to high-fidelity state transfer between nodes in a random temporal network, suggesting using temporality as a control mechanism for quantum information tasks.

Building upon these innovative approaches in quantum spatial search, our work introduces a quantum algorithm designed to discretize space amidst diverse obstacles. Our approach leverages quantum computing advancements to surpass classical algorithms' limitations in spatial discretization. Through empirical testing and validation on quantum simulators such as PennyLane, we demonstrate the superior performance of our algorithm compared to classical methods, confirming its practicality and efficiency. As the field of quantum computational geometry continues to grow, our research contributes a significant advancement, laying the groundwork for future explorations and broader applications of quantum algorithms in geometric analysis, spatial discretization, and beyond.

\section{Problem to solve}\label{sec:problem}
We consider a square enclosure \( \Omega := [0, 2^n] \times [0, 2^n] \), representing our workspace. Within this space, a set of areas \( \{\omega_i\}_{i=1}^m \) is defined, which may represent exclusion zones or areas inaccessible for various reasons, such as physical obstacles or operational constraints.

The complement of this set called the feasible area \( \Omega_f := \Omega \setminus (\bigcup_{i=1}^m \omega_i) \), constitutes the region within which we wish to perform a distribution of points. The challenge lies in generating a set of \( N \)  points such that their distribution over the feasible area \( \Omega_f \) is uniform. This issue arises in various fields, such as robotics, where the optimal placement of sensors in an environment is required, or in resource planning in computational geography.

\begin{figure}[]
\centering
    \includegraphics[width=.5\textwidth]{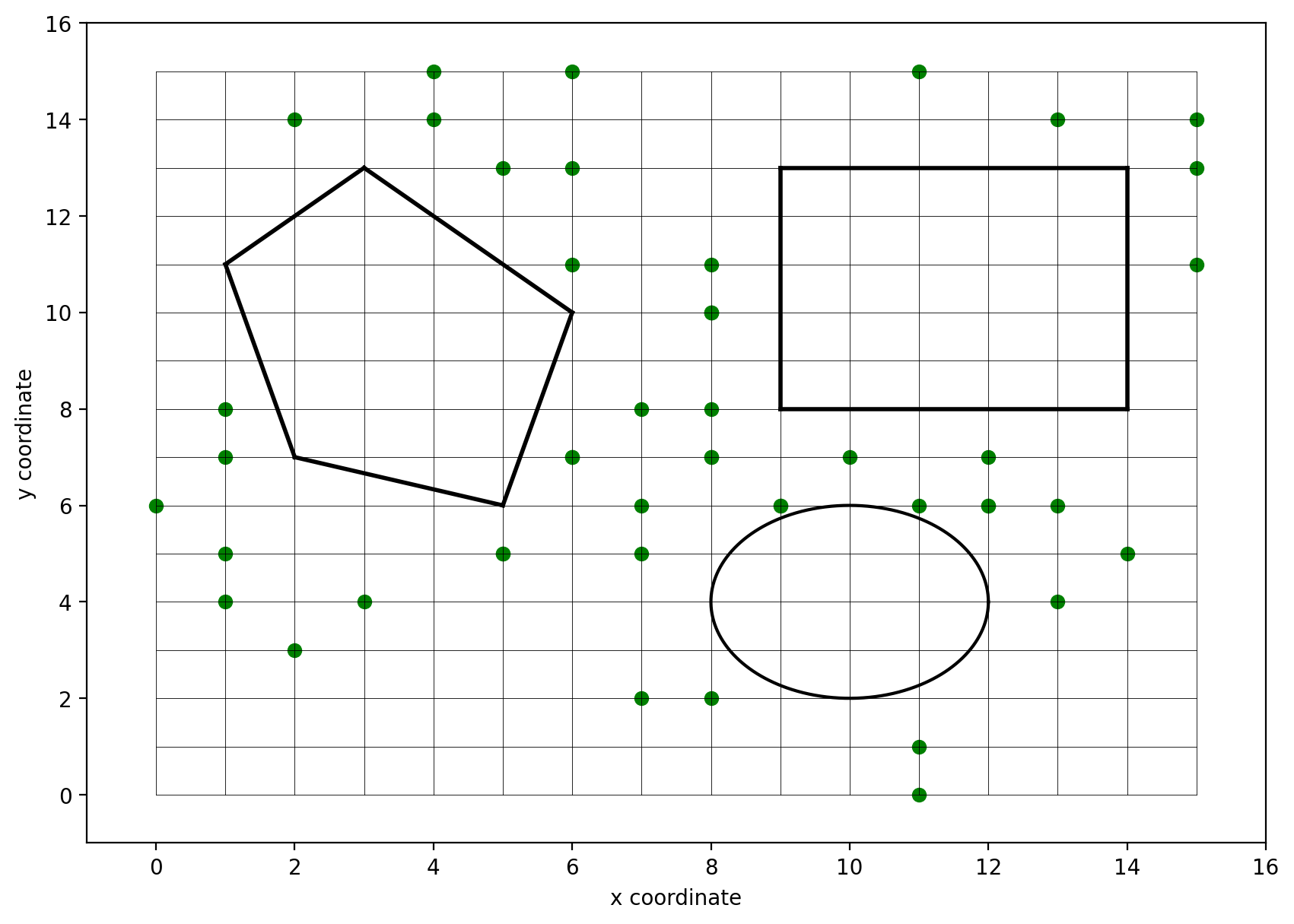}
\caption{Conceptual Illustration of the Generalized Spatial Discretization Challenge: This figure provides a 2D representation that conceptualizes the diverse geometric obstacles - rectangular, circular, and convex polygonal - encountered in spatial discretization problems. It exemplifies the broad scope of our quantum algorithm, designed to generalize and effectively handle various obstacle configurations by combining these three fundamental types.}
\label{fig:problem_to_solve}
\end{figure}

The problem focuses on generating a random sample of points that respects the constraint of uniformity in distribution over an irregularly shaped space. This implies not only considering uniformity in point density per unit area but also addressing the additional complexity imposed by the excluded areas. Consequently, we seek to develop a method that allows the generation of these points efficiently and accurately, considering the geometric constraints imposed by the set \( \{\omega_i\}_{i=1}^m \). Fig. \eqref{fig:problem_to_solve} illustrates a specific case of the problem that will be addressed in this article.

\section{Preliminaries}\label{sec:Preliminaries}

This section introduces the core concepts and definitions that will be pivotal in developing our quantum algorithmic solutions. We explore the intricate relationship between quantum computing and the geometric challenges that arise in various domains, setting the stage for a deep dive into the innovative algorithms and techniques proposed in our study.

\subsection{Notation and Definitions}
Building upon our foundational work established \cite{atchadeadelomou2023efficient}, we revisit the bifurcation of operators into two essential archetypes—\textit{Inplace} and \textit{Outplace}—which serve as the cornerstones for the execution of quantum arithmetic operations. This dichotomy not only streamlines the algorithmic architecture but also propels the frontiers of quantum computational techniques, offering a robust scaffold for developing sophisticated quantum algorithms.

\begin{definition}[Inplace Operator]
An \textit{Inplace Operator} applies a function \( f: \mathbb{N} \rightarrow \mathbb{N} \) directly onto the input qubit register, such that \( U_f\ket{x} = \ket{f(x)} \). This operator overwrites the input \( x \) with the output \( f(x) \) in the same register, adhering to the reversibility requirement of quantum computations. An example is the quantum adder by a constant \( k \), denoted as \( Add_{\text{in}}(k) \), which performs the operation \( \ket{x} \rightarrow \ket{x + k} \).
\end{definition}

\begin{definition}[Outplace Operator]
An \textit{Outplace Operator} for a function \( f: \mathbb{N} \rightarrow \mathbb{N} \), represented by the quantum operator \( U_f \), operates on two separate qubit registers. It leaves the original input register \( |x\rangle \) unaltered and outputs the function's result into an auxiliary register initially set to \( |0\rangle \), resulting in the transformation \( U_f|x\rangle|0\rangle = |x\rangle|f(x)\rangle \). This type of operator is necessary for executing non-bijective functions within quantum computing. An example is the Outplace quantum adder \( Add_{\text{out}}(k) \), which calculates \( Add_{\text{out}}(k)\ket{a}\ket{0} \rightarrow \ket{a}\ket{a+k} \), using an auxiliary register to hold the result.
\end{definition}

We now present the quantum operators pivotal to our methodology for discretizing spatial spaces in the presence of obstacles. These operators are delineated as follows:

\begin{definition}[Interval Check Operator (ICO)]
\(ICO \) is a quantum operator used to determine if a given number \( x \) is within a specific interval \( (a_1, a_2) \). It is formally defined by the action of the operator on a quantum state:
\begin{equation}
    ICO (a_1,a_2)\ket{0}\ket{x}\ket{0}\ket{0} \mapsto \ket{0}\ket{x}\ket{0}\ket{\text{s}},
    \label{eq:P_I}
\end{equation}
where \(\ket{\text{s}}\) is an ancillary qubit that indicates the result of the interval check.
\end{definition}

\begin{definition}[Rectangle Inclusion Operator (RIO)]
\( RIO \) is a quantum operator that verifies if a point $z$ with coordinates \( (x, y) \) is located within the boundaries of a defined rectangle $R_i$. Its action is represented as:
\begin{equation}
    RIO(R)\ket{x}\ket{0}\ket{y}\ket{0} \ket{0} \mapsto \ket{x}\ket{0}\ket{y}\ket{0}\ket{s},
     \label{eq:RIO}
\end{equation}
with \(\ket{\text{s}}\) being an ancillary qubit that signifies whether the point is inside the rectangle.
\end{definition}

\begin{definition}[Multiple Rectangles Inclusion Operator (MRIO)]
\( MRIO \) extends the functionality of the Rectangle Inclusion Operator (RIO) to assess the inclusion of a point in multiple rectangular regions. It is defined as follows:

\begin{equation}
    MRIO(\{R_i\}_{i=1}^m)\ket{z}\ket{0}\ket{0} \mapsto \ket{z}\ket{0}\ket{\sum_{i=1}^m \delta_i(z, R_i)},
    \label{eq:MRIO}
\end{equation}

where \(\ket{0}\) denotes an auxiliary qubit register initialized to zero, and the final qubit accumulates the count of rectangles \(R_i\) that encompass point \(z:=(x,y)\). The function \(\delta_i(z, R_i) = 1\) if point \(z\) is within rectangle \(R_i\), and \(\delta_i(z, R_i) = 0\) otherwise. This operator facilitates the simultaneous evaluation of point inclusion across multiple geometric regions.
\end{definition}

\begin{definition}[Quantum Threshold Comparator (\( \hat{T} \))]
\( \hat{T} \) is a quantum operator used to determine whether a given quantum state \( \ket{x} \) representing a value \( x \) is greater than or equal to a certain threshold \( c \). It encodes this comparison as a binary outcome in an ancillary qubit \( \ket{\alpha} \), which can be used in further quantum computations. The operator is defined as:
\begin{equation}
    \hat{T}\ket{x}\ket{0} = \ket{x}\ket{\alpha},
    \label{eq:operator_T}
\end{equation}
where \( \ket{\alpha} \) is an ancillary qubit that indicates the result of the comparison: it is set to \( \ket{0} \) if \( x \geq c \), and to \( \ket{1} \) if \( x < c \).
\end{definition}

\begin{definition}[Quantum Negation Operator (\( \hat{N} \))]
\( \hat{N} \), also known as the quantum bitwise complement operator, negates a binary number represented in a quantum register. For a number \( x \) less than a predefined modulus \( 2^n \), the operator \( \hat{N} \) computes its negation modulo \( 2^n \), effectively calculating \( 2^n - x \). This is particularly useful in quantum arithmetic circuits where the inversion of binary values is required. The action of \( \hat{N} \) is given by:
\begin{equation}
    \hat{N}\ket{0}\ket{x} = \ket{2^n - x},
    \label{eq:operator_N}
\end{equation}
where the first quantum register is typically an ancillary qubit initialized to \( \ket{0} \), and \( \ket{x} \) is the quantum register holding the value to be negated.
\end{definition}

\begin{definition}[Outplace Quantum Adder Operator (\( \text{Add}_{out}(k) \))]
\( \text{Add}_{out}(k) \) is a quantum operator that adds a classical integer \( k \) to the quantum state representing an integer \( x \) in a quantum register. It acts on a superposition of states and applies the addition coherently to each element in the superposition, maintaining quantum parallelism. The operator is defined as follows:
\begin{equation}
    \text{Add}_{out}(k)\ket{x}\ket{0}^{\otimes m} = \ket{x + k},
\end{equation}
where \( \ket{x} \) is an \( n \)-qubit quantum register encoding the integer \( x \), and \( \ket{0}^{\otimes m} \) are \( m \) ancillary qubits provided to facilitate the addition, with \( m \) typically determined by the space requirements of the addition operation. The result \( \ket{x + k} \) represents the new state of the quantum register after the addition of \( k \).
\end{definition}

\begin{definition}[Inplace Quantum Adder (\( Add_{\text{in}} \))]
\( Add_{\text{in}} \) is a quantum operator that adds two quantum registers in an \textit{Inplace} manner. Given two quantum registers \( \ket{a} \) and \( \ket{b} \), the operator \( Add_{\text{in}} \) applies the transformation \( \ket{a}\ket{b} \rightarrow \ket{a}\ket{a+b} \). This operator is designed to preserve the value of the first register \( \ket{a} \) while adding it to the second register \( \ket{b} \), updating \( \ket{b} \) with the result of the addition.
    \begin{equation}
    Add_{\text{in}}\ket{a}\ket{b} \rightarrow \ket{a}\ket{a+b}.
    \label{eq:adder_in_q_q}
\end{equation}
\end{definition}

\begin{definition}[Inplace Quantum Square Adder Operator (\( \text{AddSqr}_{in} \))]
\( \text{AddSqr}_{in} \) is a specialized quantum operator used for adding the square of the value encoded in one qubit register to another. Formally, for qubit registers \( \ket{x} \) and \( \ket{y} \), the action of \( \text{AddSqr}_{in} \) is defined as:

\begin{equation}
    \label{eq:Addsqr}
    \text{AddSqr}_{in}\ket{x}\ket{y} = \ket{x}\ket{y + x^2},
\end{equation}

where, \( \ket{x} \) is the input register holding the value to be squared, and \( \ket{y} \) is the target register where the squared value is added. 
\end{definition}

\begin{definition}[Outplace Quantum Multiplication Operator \( \text{Mult}_{out}(k) \)]
\( \text{Mult}_{out}(k) \) is a theoretical construct representing a class of quantum operations that, given a classical multiplier \( k \) and a quantum register \( \ket{x} \) encoding a quantum state, produces a new quantum state where the quantum register is entangled with an ancillary register to represent the product \( kx \). The operator is defined as:
\begin{equation}
    \text{Mult}_{out}(k)\ket{x}\ket{0} := \ket{x}\ket{kx},
    \label{eq:mult_operator}
\end{equation}
where \( \ket{x} \) is the original quantum state, \( \ket{0} \) is the initial state of the ancillary quantum register, and \( \ket{kx} \) represents the quantum state after the application of the multiplication operation.
\end{definition}

\begin{definition}[Outplace Quantum Absolute Difference Circuit ($\text{AbsDiff}_{out}$)]
Let \( \ket{x} \) be a quantum state encoding the binary representation of a number \( x \), and let \( k \) be a classical number. The \emph{Quantum Absolute Difference Circuit} is a quantum operation that transforms an initial state \( \ket{x}\ket{0}\ket{0}\ket{0} \) into a final state \( \ket{x}\ket{0}\ket{0}\ket{|x - k|} \), where \( \ket{|x - k|} \) encodes the absolute difference between \( x \) and \( k \). This operation is represented as:

\begin{equation}
    \text{AbsDiff}_{out}(k)\ket{x}\ket{0}\ket{0}\ket{0} \rightarrow \ket{x}\ket{0}\ket{0}\ket{|x - k|}.
    \label{eq:absDiff}
\end{equation}

\end{definition}

These operators will be used to construct the quantum circuits necessary for efficiently discretizing spaces with obstacles. They will be employed within Grover's algorithm to enhance the process of finding a feasible region within a discretized space.

\subsection{Quantum Discretization Across Geometric Landscapes}
Quantum algorithmic geometry often encounters various obstacles, especially in discretizing spatial environments. Our approach classifies these into three primary scenarios—rectangular, circular, and convex polygonal obstacles—each demanding specific quantum computational strategies:

\begin{itemize}
    \item \textit{Rectangular Obstacles:} These are common in engineered environments, presenting orthogonal spatial challenges requiring specialized quantum solutions.
    \item \textit{Circular Obstacles:} Frequently encountered in applications where radial symmetry is a factor, they require custom quantum discretization techniques.
    \item \textit{Convex Polygonal Obstacles:} These cover a wide range of geometric shapes, crucial in graphical simulations and design, where quantum computations must be precise.
\end{itemize}

Our framework constructs quantum oracles for each obstacle type, which are vital components of our algorithmic solution. This process involves creating specific quantum circuits and leveraging Grover's algorithm \cite{grover1996fast} to enhance the efficiency and precision of the discretization process.

In dealing with \textit{Rectangular Obstacles}, our approach includes circuits designed for point inclusion verification, utilizing operators for boundary checks and summation. In the \textit{Circular Obstacle} scenario, we employ circuits that integrate general addition, absolute value computation, and distance measurement operations. When addressing \textit{Convex Polygonal Obstacles}, the focus shifts to integer arithmetic within quantum environments, which involves using operators for scalar arithmetic and vector cross-products.

These detailed circuit designs, tailored for specific obstacle types, are adaptable for environments with multiple shapes. The upcoming sections will delve deeper into these concepts, exploring the algorithmic nuances and the strategic development of quantum gates and circuits. This is essential for achieving our discretization goals within these varied geometric landscapes.

Our methodology addresses the general case by considering these three obstacle types: rectangular, circular, and convex polygons. This comprehensive approach lays the foundation for effectively managing and representing obstacles in diverse spatial discretization applications.

\section{Implementation}\label{sec:IMPLEMENT}

\subsection{Spatial Discretization Through a Quantum Algorithm}

\begin{figure}[]
\centering
    \includegraphics[width=.5\textwidth]{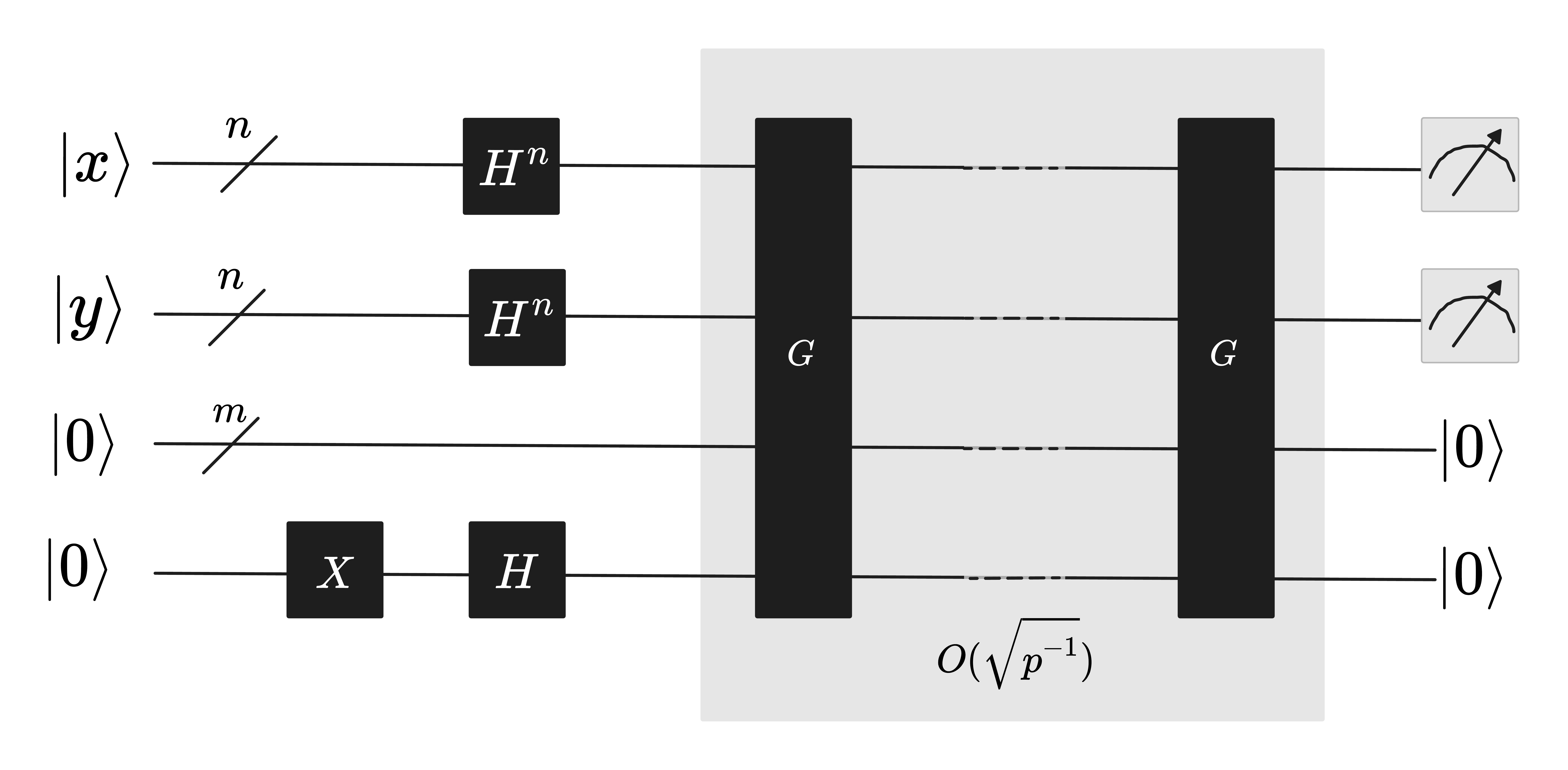}
\caption{Quantum circuit utilizing Grover's algorithm to efficiently discretize the feasible region \( \Omega_f \). The circuit is repeated \( N \) times to achieve the desired discretization.}
\label{fig:quantum_circuit}
\end{figure}

Suppose we have access to a function, which we will call \( O: \mathbb{R}^2 \rightarrow \{0,1\} \), that determines if a point $z:=(x,y)$ is within the feasible area. Let \( A_f \) be the area of the feasible region and \( A_t \) the total area; then the probability of randomly choosing a point in the viable area is \( p := \frac{A_f}{A_t} \). If this probability is unknown, it can be estimated simply using an algorithm like Monte Carlo \cite{metropolis1909beginning, grover1996fast,hansmann1999new}. Suppose we want to discretize the plane into \( N \) points. Using classical computing techniques, the general strategy consists randomly selects a point from space and checks whether the function is in the feasible area with the function \( O \). The expected number of times we would need to use the function \( O \) following this method is \( \frac{N}{p} \). However, this technique can be improved with quantum computing. By applying Grover's algorithm \cite{grover1996fast}, we can reduce this number. If \( p \) is the probability of choosing a point in the feasible area, Grover's algorithm theory tells us that by applying Grover's iterator \( O(\frac{1}{\sqrt{p}}) \) times, we can find a point in the feasible area with a probability close to 1. This approach reduces the number of times we need to apply the oracle to \( O(\frac{N}{\sqrt{p}}) \) (please refer to the mathematical demonstration in Appendix \eqref{sec:appendix}), thus achieving a quadratic improvement over the classical case. Fig. \eqref{fig:quantum_circuit} schematizes the theory we have just developed.

By repeating this circuit \( N \) times, we acquire \( N \) points that make up the desired discretization of the region. One of the primary challenges when applying Grover's algorithm is constructing the quantum oracle. In the following sections, we offer detailed examples for various scenarios and explain how to implement the oracle for each case.

\subsection{Oracle for Rectangular Obstacle Problems}

\begin{figure}[]
\centering
    \includegraphics[width=.3\textwidth]{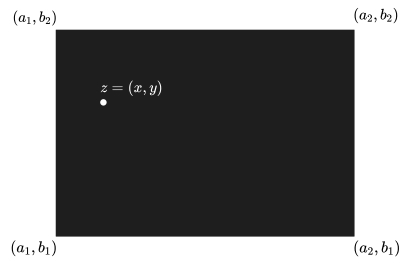}
\caption{This figure illustrates the rectangular obstacle and given a point $z=(x,y)$ falls within a rectangular boundaries. Each rectangle is defined by its corner coordinates: $(a_1, b_1)$, $(a_1, b_2)$, $(a_2, b_1)$, and $(a_2, b_2)$. }
\label{fig:rectangular_obs}
\end{figure}

This section details the construction of Grover's oracle tailored for environments featuring rectangular obstacles. Fig. \eqref{fig:rectangular_obs} presents the rectangular obstacle with all the variables for constructing the oracle. 

We consider a point \( z \) situated within a two-dimensional plane, described by the coordinate pair \( (x, y) \). The coordinates \( x \) and \( y \) specify the point's location along the horizontal and vertical axes. Our objective is to ascertain whether the point \( z \), represented quantumly as \( \ket{x_1,\ldots,x_n}\ket{y_1,\ldots,y_n} \), lies within a specified rectangular region. This region is defined by the Cartesian product of intervals \( (a_1, a_2) \) and \( (b_1, b_2) \). To achieve this, we deploy a set of three distinct quantum operators:

\begin{itemize}
    \item \( ICO \): Checks if \( a_1 \leq x \leq a_2 \). See Equation \eqref{eq:P_I}.
    \item \( RIO \): Verifies if \( (x, y) \) is inside the rectangle \( R \). See Equation \eqref{eq:RIO}.
    \item \( MRIO \): Determines if a point \( (z) \) lies within any given rectangle. See Equation \eqref{eq:MRIO}.
\end{itemize}

\begin{figure}[]
\centering
    \includegraphics[width=.5\textwidth]{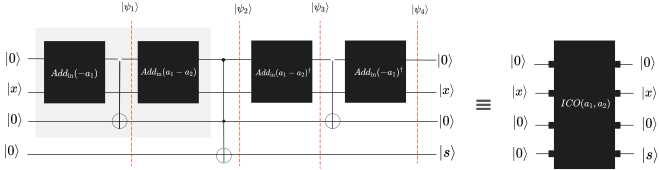}
\caption{Quantum circuit for the \( ICO \) operator, checking if a number \( x \) is within the interval \( (a_1, a_2) \). The solution remains in the $s$ register. }
\label{fig:pi_circuit}
\end{figure}

We present the circuit \( ICO \) which verifies if a number \( x \in \mathbb{N} \) is within the interval \( (a_1, a_2) \) (see Figure \ref{fig:pi_circuit}).

\begin{figure}[]
\centering
    \includegraphics[width=.5\textwidth]{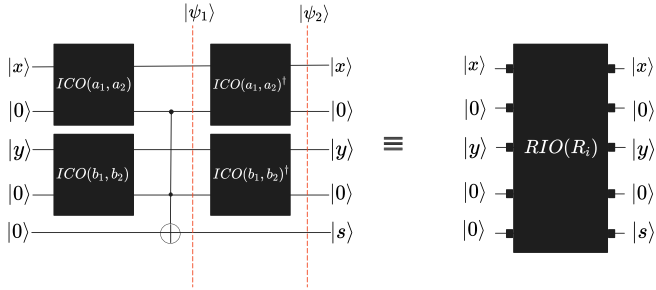}
\caption{The Quantum circuit for the $RIO$ operator confirms whether a point $(x,y)$ lies within the i-th rectangle, denoted as $R_i$.}
\label{fig:pr_circuit}
\end{figure}

The \( ICO \) operator uses an auxiliary qubit to store the result and requires three additional qubits. The construction of the \( RIO \) operator is depicted in Figure \ref{fig:pr_circuit}.

\begin{figure}[]
\centering
    \includegraphics[width=.5\textwidth]{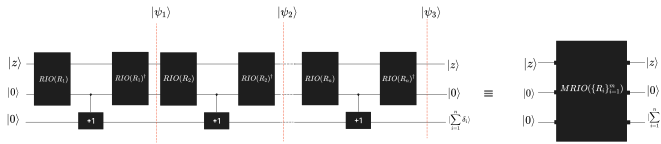}
\caption{Quantum circuit for the \( MRIO \) operator, verifying if a point \( z \) is inside a rectangle $\{R_i\}_{i=1}^m$.}
\label{fig:mrio}
\end{figure}

The \textit{MRIO} operator \ref{fig:mrio} requires using three auxiliary qubits to implement $ICO$, two more auxiliary qubits for storing its outcomes, and an additional qubit for the final result, amounting to six qubits. Once $RIO$ is constructed, developing $MRIO$ for multiple rectangles becomes straightforward. It simply involves applying $RIO$ for each of the available rectangles. As defined in the construction of the $RIO$ operator, each of these operators takes the set $R_i := (a_{i1}, a_{i2}, b_{i1}, b_{i2})$ as parameters, which define the rectangle $(a_{i1}, a_{i2}) \times (b_{i1}, b_{i2})$. Let us recall Eq. \eqref{eq:RIO}, and from it, let us explain its evolution through Fig. \eqref{fig:pr_circuit} as follows:
\begin{itemize}
\item $\psi_1 := \ket{x}\ket{s_1}\ket{y}\ket{s_2}\ket{s}$: After applying the $RIO$ operator to the $x$ and $y$ registers, the auxiliary qubits are employed to determine if these values fall within the specified intervals.
\item $\psi_2 := \ket{x}\ket{0}\ket{y}\ket{0}\ket{s}$: Upon completion of the algorithm, the last qubit stores the outcome, indicating whether the point resides within the desired interval.
\end{itemize}

Let us recall the definition of $MRIO$ \eqref{eq:MRIO} and explain the algorithm's intermediate steps. We already know that $\delta_i = 1$ when the point $z$ belongs to the rectangle $R_i$, and $\delta_i = 0$ otherwise, then the steps are as follows:
\begin{itemize}
    \item $\psi_1 := \ket{z}\ket{0}\ket{\delta_1}$. Initially, we check if the point $z$ belongs to the rectangle $R_1$. If it does, we add one to the last register.
    \item $\psi_2 := \ket{z}\ket{0}\ket{\delta_1 + \delta_2}$. We repeat the procedure for the second rectangle.
    \item $\psi_3 := \ket{z}\ket{0}\ket{\sum_{i=1}^n \delta_i}$. After checking if the point $z$ belongs to each of the rectangles, the last register will hold the value of the number of rectangles to which the point $z$ belongs.
\end{itemize}

\subsection{Oracle for Circular Obstacle Problems}
\label{sec:oracle_circular}
\begin{figure}[!ht]

\centering
    \includegraphics[width=.3\textwidth]{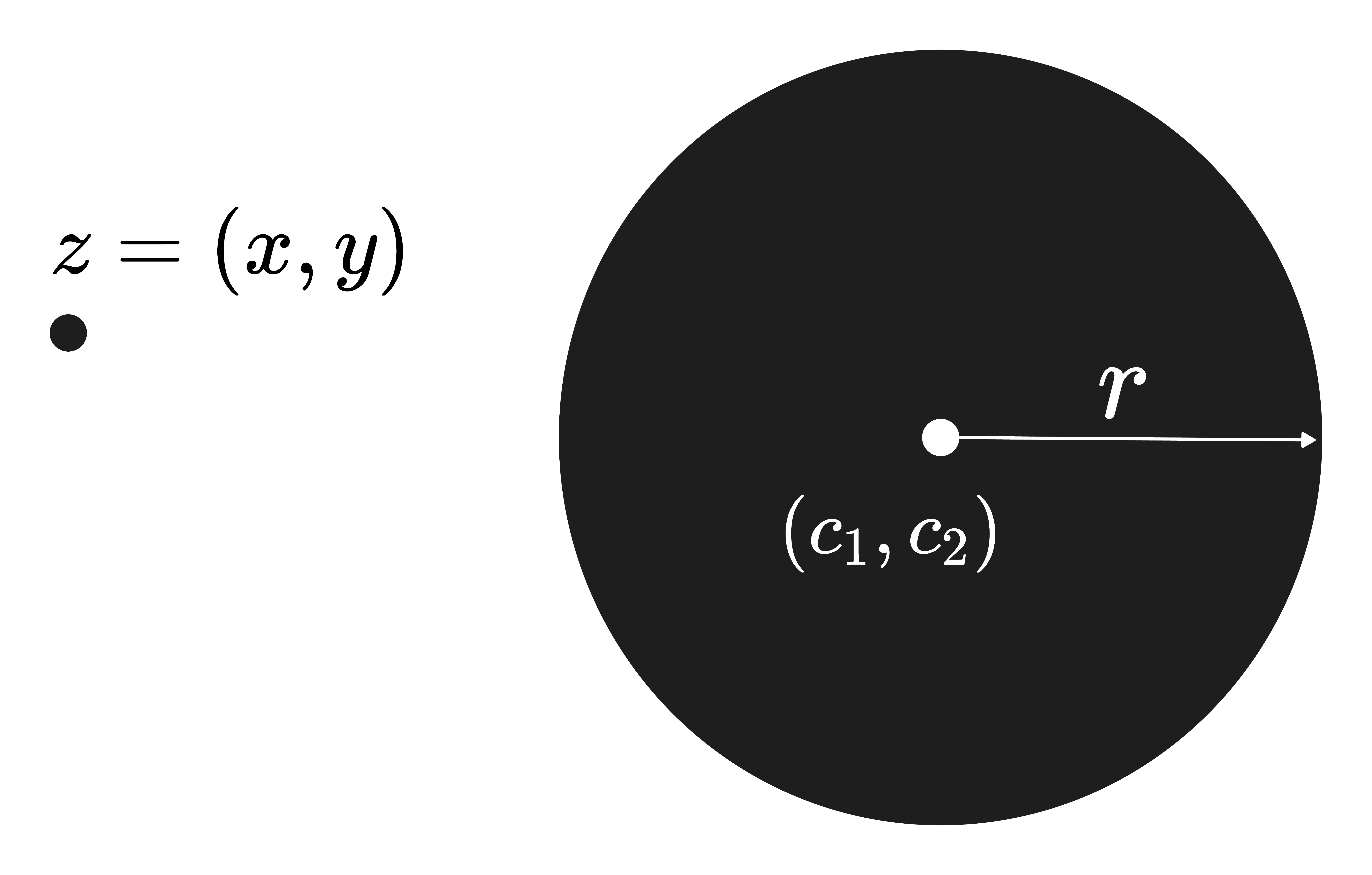}
\caption{This figure illustrates the circular obstacle with a given point $z=(x,y)$. A center has two coordinates ($c_1, c_2$) and a radius ($r$) that define each circular obstacle, and the circuit evaluates whether the point $z=(x,y)$ lies outside these circular regions.}
\label{fig:circular_obs}
\end{figure}

 Our quantum oracle is crafted to discern whether a point \( (x,y) \) falls outside the boundary of a circular obstacle, defined by centers \( \{c_i\}_{i=1}^m \) and radii \( \{r_i\}_{i=1}^m \). Fig. \eqref{fig:circular_obs} presents the circular obstacle with its variables.

The exclusion condition for a point from a circular obstacle's area is given by:
\begin{equation}
    r_i^2 \leq (x-c_{1i})^2 + (y-c_{2i})^2.
    \label{eq:circle_inequality}
\end{equation}

To address the non-bijectivity challenge of the quadratic function $h(x):=(x-c_1)^2$ in quantum computation, we employ the \textit{Quantum Threshold Comparator} (\( \hat{T} \)) \eqref{eq:operator_T} and the \textit{Quantum Negation Operator} (\( \hat{N} \)) \eqref{eq:operator_N} which are central to computing the Euclidean distance following Equation \ref{eq:circle_inequality}.

But, before computing the Euclidean distance, we must construct all the ingredients to reach this point. First, we must construct the \textit{AbsDiff} operator defined in Eq. \eqref{eq:absDiff}.

The binary variable \( \alpha \) \eqref{eq:alpha_definition}, which indicates whether \( x \) is greater or less than \( c_1 \), is determined as follows:

\begin{equation}
\alpha := 
    \begin{cases} 
        0 & \text{if } c_1 \leq x, \\ 
        1 & \text{if } x < c_1. 
    \end{cases}    
    \label{eq:alpha_definition}
\end{equation}

Using \( \alpha \), the function \( f(x) \) \eqref{eq:quantum_function_f} for computing the absolute difference needed for the Euclidean distance is defined by:

\begin{equation}
    f(x) = (1-\alpha)(x - c_1) + \alpha(-x + c_1).
    \label{eq:quantum_function_f}
\end{equation}



When \( x < c_1 \), the \textit{Quantum Negation Operator} \( \hat{N} \) \eqref{eq:operator_N} is used to calculate \( -x \).

The quantum circuit implementing Equation \ref{eq:quantum_function_f} is illustrated in Figure \ref{fig:absolute_value_circuit}, which depicts the temporal evolution of the quantum state and necessitates the following operators:

\begin{figure*}[]
\centering
    \includegraphics[width=1\textwidth]{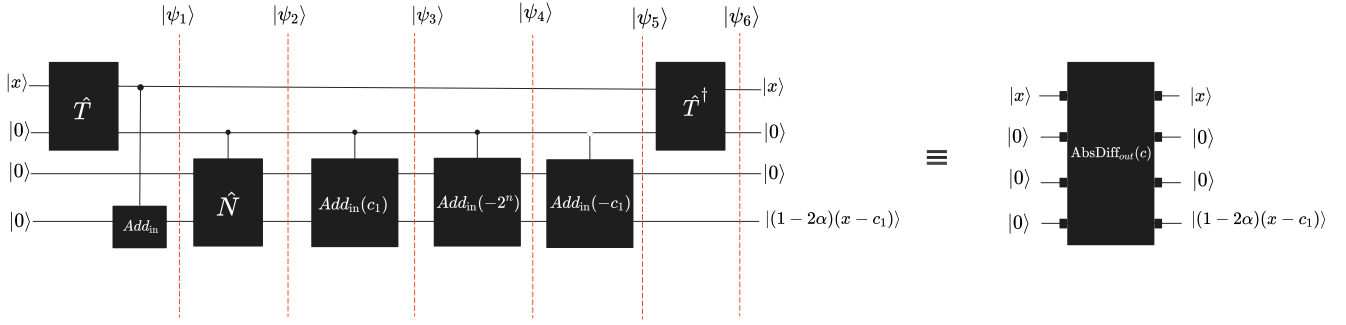}
\caption{Quantum State Evolution for the Circular Obstacle Oracle. This sequence depicts the intricate process of computing absolute distances and resetting the system: $\psi_1$ initiates with the encoding of primary values; $\psi_2$ and $\psi_3$ engage in conditional operations to compute the absolute difference $|x - c_1|$, where $c_1$ represents the center of the circular obstacle; $\psi_4$ through $\psi_6$ culminate the computation process and systematically reset the ancillary qubits for future computations, ensuring the efficient and reusable nature of the quantum oracle.}
\label{fig:absolute_value_circuit}
\end{figure*}

The evolution of the circuit's state is detailed step-by-step, elucidating the critical role of \( \hat{T} \) and \( \hat{N} \) in facilitating the quantum computation of the Euclidean distance, thus validating our approach for solving circular obstacle problems within a quantum context.

This oracle exemplifies the application of quantum computational techniques to classical geometric problems and showcases the potential for significant advancements in quantum navigation algorithms.

Reordering \eqref{eq:quantum_function_f}, we get to Eq. \eqref{eq:quantum_function_f_ext}

\begin{equation}
    f(x)=(1-2\alpha)(x - c_1).
    \label{eq:quantum_function_f_ext}
\end{equation}

Once we have this operator, we can proceed as follows.

Let us examine how the circuit's state from Fig. \eqref{fig:absolute_value_circuit} evolves at each labeled step.

\begin{itemize}
\item $\psi_1 := \ket{x}\ket{\alpha}\ket{0}\ket{x}$: Initially, the state $\psi_1$ represents the system after the application of operator $\hat{T}$, which encodes the value of $\alpha$ based on the comparison of $x$ and $c_1$. The ancillary qubits are in their initialized states, and the register holds the value of $x$.

\item $\psi_2 := \ket{x}\ket{\alpha}\ket{(1-\alpha)x + \alpha(2^n - x)}$: At this juncture, the circuit has engaged in a conditional operation based on the value of $\alpha$. The state remains unchanged if $\alpha$ is 0, indicating $x \geq c_1$. Otherwise, the system negates $x$, calculating $2^n - x$.

\item $\psi_3 := \ket{x}\ket{\alpha}\ket{(1-\alpha)x + \alpha(2^n - x + c_1)}$: The state $\psi_3$ reflects the addition of $c_1$ to the negated value of $x$ when $\alpha$ is 1. This is part of the process to compute the absolute difference $|x - c_1|$.

\item $\psi_4 := \ket{x}\ket{\alpha}\ket{(1-\alpha)x + \alpha(c_1 - x)}$: Here, the system corrects the earlier negation by subtracting $2^n$ when $\alpha$ is 1, completing the computation of $|x - c_1|$.

\item $\psi_5 := \ket{x}\ket{\alpha}\ket{0}\ket{(1-\alpha)(x - c_1) + \alpha(c_1 - x)}$: At this stage, $\psi_5$ shows the ancillary qubits being reset, leaving the final qubit register with the absolute difference computed, ready for further operations.

\item $\psi_6 := \ket{x}\ket{0}\ket{0}\ket{(1-2\alpha)(x - c_1)}$: Finally, we reapply the operator $\hat{T}^\dagger$ to not compute the value of $\alpha$, resetting the auxiliary qubit, and leaving the system with only the calculated absolute difference. This step is crucial for restoring the system to a state where the additional qubits can be reused for subsequent computations without carrying over any information from the previous operation.

\end{itemize}

Once we know how to calculate $h(x) = |x - c_1|$, the rest of the operations are straightforward.

\begin{figure}[]
\centering
    \includegraphics[width=.5\textwidth]{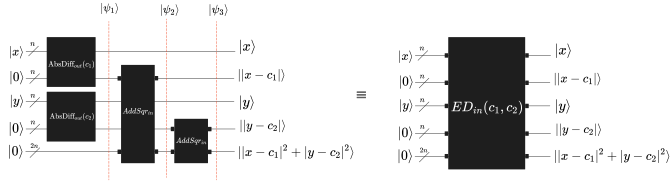}
\caption{Stages of quantum state evolution in Euclidean distance calculation. Starting with $\psi_1 := \ket{x, |x - c_1|, y, |y - c_2|, 0}$, the register is initialized and prepared for encoding the Euclidean distance. The state $\psi_2 := \ket{x}\ket{|x - c_1|}\ket{y}\ket{|y - c_2|}\ket{|x - c_1|^2}$ advances the calculation by squaring the encoded distances, storing essential terms for proximity evaluation. The final state $\psi_3 := \ket{x}\ket{|x-c_1|}\ket{y}\ket{|y-c_2|} \ket{|x-c_1|^2 + |y-c_2|^2}$ reflects the register after uncomputation, resetting ancillary qubits and leaving the computed squared distance in the last qubit.}
\label{fig:Circular_Oracle}
\end{figure}

The quantum state's evolution in the circular obstacle oracle unfolds through precise arithmetic operations. We detail the progression from $\psi_1$ to $\psi_3$, revealing the complex computational evolution depicted in Fig. \eqref{fig:Circular_Oracle}:

\begin{itemize}
    \item $\psi_1 := \ket{x, |x-c_1|, y, |y-c_2|, 0 }$ represents the quantum register initialized with the coordinate values and ancillary qubits prepared to encode the Euclidean distance computation's intermediate steps. This state captures the preparation phase where the groundwork for distance calculations is laid.
    
    \item The system progresses to the execution of the squared distance computation transitioning to $\psi_2 := \ket{x}\ket{\mid x-c_1 \mid}\ket{y}\ket{\mid y-c_2 \mid}\ket{\mid x-c_1 \mid ^2}$,  At this juncture, the ancillary qubits store the squared terms of the Euclidean distance, a pivotal step that converges towards determining point-to-obstacle proximity.
    
    \item Lastly, unveils the quantum register post 'uncomputation' $\psi_3 := \ket{x}\ket{|x-c_1|}\ket{y}\ket{|y-c_2|} \ket{|x-c_1|^2 + |y-c_2|^2}$. This state demonstrates the quantum circuit's reversibility, where ancillary qubits are reset, and the register is restored to a near-initial state—without the last qubit, which now holds the essential computed value for the discretization task.
\end{itemize}

The quantum state evolution within the oracle reflects a work of quantum arithmetic, ensuring the reversibility of computations—fundamental to quantum mechanics. This procedural integrity is vital for precisely determining spatial relationships between a point and circular obstacles, epitomizing the remarkable efficiency of quantum computation.

The requisite qubit allocation for the oracle is as follows:
\begin{itemize}
    \item A total of $2n$ qubits are allocated for encoding the coordinate variables $x$ and $y$.
    \item An ensemble of $2n + 1$ qubits is dedicated to calculating the squared Euclidean distance, specifically $|x - c_1|^2 + |y - c_2|^2$.
    \item An additional $n + 1$ auxiliary qubits are employed to assist in the intermediate computations, namely $|x - c_1|$ and $|y - c_2|$.
\end{itemize}

Consequently, implementing this quantum circuit requires $5n + 1$ qubits, optimizing the oracle construction to a more efficient requirement of $3n + 1$ qubits, exemplifying optimal resource utilization within quantum computational frameworks.

\subsection{Oracle for Convex Polygon Problems}
\begin{figure}[!h]
\centering
    \includegraphics[width=.3\textwidth]{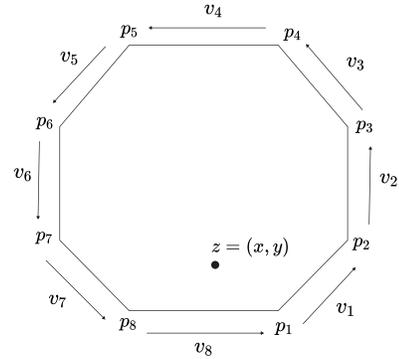}
\caption{This figure illustrates a convex polygonal obstacle to ascertain whether a point $z=(x,y)$ is located within the confines of an eight-sided polygon. The polygon is defined by a sequence of eight vertices $\{v_1, v_2, \ldots, v_8\}$ and corresponding edge vectors $\{p_1, p_2, \ldots, p_8\}$. The circuit evaluates the spatial relationship between point $z$ and each polygon edge to determine whether $z$ falls inside the polygonal boundary.}
\label{fig:polygon_obs}
\end{figure}

In cases where obstacles are generalized convex polygons, we aim to verify that each point lies within the polygon. Should the polygon be concave, a subdivision technique into convex polygons is applicable. Fig. \eqref{fig:polygon_obs} presents a convex polygon obstacle and its variables.
 We consider the following scenario: a convex polygon whose \( m \) sides are defined by a set of vertices \( \{p_i\}_{i = 1}^m \) and vectors \( \{v_i\}_{i=1}^m \). To ascertain whether a point \( (x,y) \) falls to the left or right side of the line $\{p_i + \lambda v_i: \lambda \in \mathbb{R}\}$, where the sign of the cross products \( (p_i-z)\land v_i \) needs to be calculated.

For the calculations just described, it is necessary to employ integer arithmetic \cite{csahin2020quantum}.

\subsubsection{Construction of the Oracle}
Consider a polygon delineated by a sequence of vertices \(\{p_i\}_{i=1}^M\). We construct vectors \(v_i := \overline{p_ip_{i+1}}\), where each \(v_i\) connects consecutive vertices \(p_i\) and \(p_{i+1}\). Collectively, these vertices and vectors give rise to the series of defining lines \(\{R_i\}_{i=1}^M\). Inclusion of a point \(x\) within the polygon is affirmed if all the vector cross-products \((x-p_i) \times v_i\) consistently share the same sign, indicating that \(x\) consistently lie on the same side of each line \(R_i\) (see Fig. \eqref{fig:polygon_obs}). To facilitate this verification, we introduce a quantum circuit responsible for calculating the vector cross-products, integrating both quantum and classical vectors, as depicted in Fig. \eqref{fig:op_prod_vectorial}.

\subsubsection{Cross-Product Computation in Quantum Circuits}

The crux of evaluating a point's position relative to a polygonal boundary lies in computing the vector cross-product. This necessitates a quantum circuit (see Fig. \ref{fig:op_prod_vectorial}) that operationalizes the following transformation:

\begin{equation}
\begin{split}
    VCP(v_1, v_2)\ket{S_x}\ket{x}\ket{S_y}\ket{y}\ket{0} \mapsto \\
    \ket{S_x}\ket{x}\ket{S_y}\ket{y}\ket{xv_2 - yv_1},
\end{split}
\label{eq:vector_cross_product}
\end{equation}

where \( VCP(v_1, v_2) \) is a unitary operator that maps the inputs to the cross-product of two-dimensional vectors, a fundamental operation in determining the positional relation of a point to the edges of a polygon.

To facilitate this calculation, our approach integrates integer arithmetic within the quantum framework, utilizing two complementing notations: absolute value-sign notation for efficient multiplication and two's complement notation for straightforward addition, as introduced by Baugh and Wooley \cite{baugh1973two}. These notations enable the quantum circuit to perform arithmetic operations with enhanced simplicity and precision. At the heart of this arithmetic process is the multiplication operator \( Mult \), defined for natural numbers \( x \) in Equation \eqref{eq:mult_operator}. Through the application of \( Mult \), our circuit (\ref{fig:op_prod_vectorial}) achieves the cross-product, enabling the quantum evaluation of point inclusion in polygonal regions with unprecedented efficiency. Then, the resultant quantum circuit is as follows:

\begin{figure*}[ht]
\centering
    \includegraphics[width=1\textwidth]{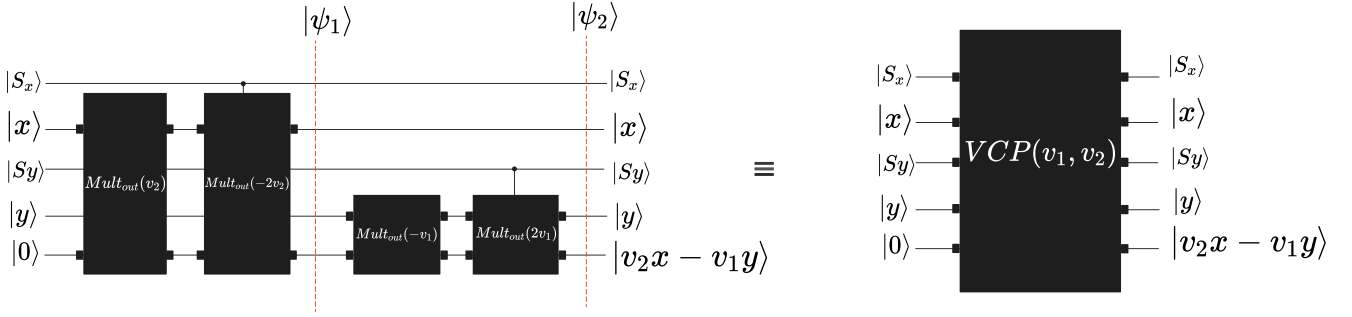}
\caption{Quantum circuit for the Cross-Product Operator as defined in Equation \ref{eq:vector_cross_product}. The initial state $\psi_1$ reflects the computation of the term $v_1x$, setting up for the cross-product. The subsequent state $\psi_2$ captures the completion of the cross-product with the result $xv_2 - yv_1$, illustrating the critical step in vector operations within quantum computing.}
\label{fig:op_prod_vectorial}
\end{figure*}


\begin{itemize}
    \item The quantum state is given by $\psi_1 = \ket{s_x}\ket{x}\ket{s_y}\ket{y}\ket{v_1x}$. After applying the first part of the circuit, we obtain the desired calculation.
    \item The circuit output is observed on $\psi_2 = \ket{s_x}\ket{x}\ket{s_y}\ket{y}\ket{v_2x - v_1y}$.
\end{itemize}

\begin{figure*}[]
\centering
    \includegraphics[width=1\textwidth]{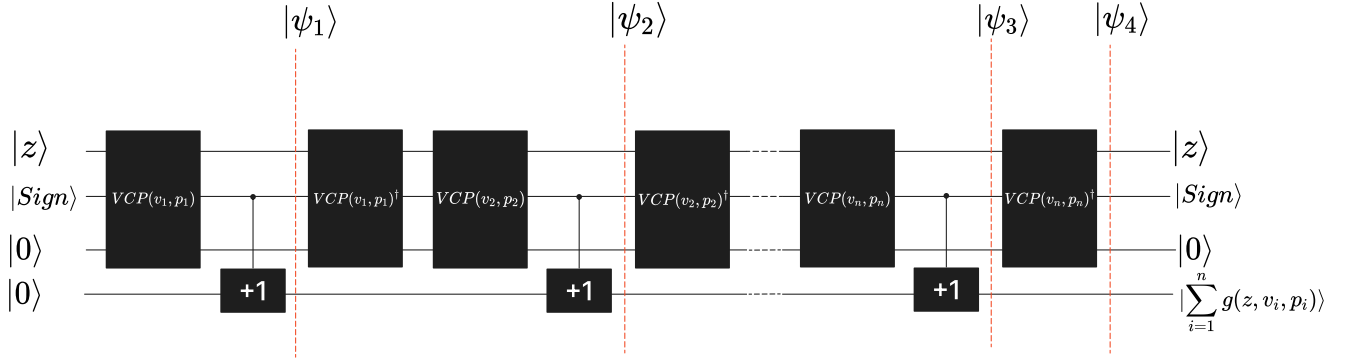}
\caption{Quantum circuit for verifying a point's inclusion within a polygon using the cross-product function \( f \) by iteratively computing and resetting vector cross-products for each polygon edge, with controlled operations indicating the point's position relative to the polygon's boundaries.}
\label{fig:polygon_oracle_circuit}
\end{figure*}

Recalling that, we aim to verify if a point \( z:=(x,y) \) is inside the polygon, let $M$ be the number of sides of the polygon and let $f$ be the function defined by $f(v_i, p_i)\ket{z}\ket{0} \rightarrow \ket{z}\ket{(z-p) \times v_i}$. A point lies within the polygon if the sign of all the cross-products $(z-p_i) \times v_i$ is positive for every $i$. We define the function $g(z, v_i, p_i)$, which equals $0$ if the point $z$ is on the correct side of the line $\{p_i + \lambda v_i: \lambda \in \mathbb{R}\}$ and $1$ if it is on the incorrect side. 
Hence, we can construct the circuit \eqref{fig:polygon_oracle_circuit} accordingly.

Each step in the circuit and its impact on the quantum state is detailed, illustrating the process for determining the point's inclusion within the polygonal area as depicted in Figure \ref{fig:polygon_oracle_circuit}. 

We now consider the operations necessary to determine whether a point is within a polygon. Let us represent a polygon as a sequence of vertices \( \{v_i\}_{i=1}^n \) such that the sides of the polygon are formed by the segments \( e_i := v_i \to v_{i+1} \) for \( i \in \{1,...,n\} \) (here, \( v_1 = v_{n+1} \)). The vertices should be ordered so that the edges \( e_i \) are correctly oriented, meaning that the polygon's interior remains to the left of the edge being traversed. Once the polygons are defined in this manner, to check whether a point \( p \) belongs to the interior of the polygon, it must satisfy \( (p-v_i) \land e_i > 0 \) for all \( i \). In other words, the point \( p \) must lie to the left of each edge forming the sides. We then consider the operator \( f \) defined as follows:
\begin{equation}
    f(v_i, e_i)\ket{z} = \ket{(z-v_i) \land e_i}
\end{equation}
To devise the function $f$, it is essential to utilize the $Add_{in}$ operator, adapted to perform subtraction operations (refer to Eq. \eqref{eq:adder_in_q_q}). Additionally, the $VCP$ (Vector Cross Product) operator is employed for executing vector cross-product calculations (as delineated in Eq. \eqref{eq:vector_cross_product}). Consequently, a point is determined to be inside the polygon if the condition \( \sum_{i=1}^n f(v_i, e_i)\ket{z} = 0 \) is satisfied. Figure \eqref{fig:polygon_oracle_circuit} presents the quantum circuit designed to ascertain whether a point \( z \) resides within the confines of a polygon delineated by the vertices \( \{v_i\}_{i=1}^n \).

Let us examine the state of the circuit at each intermediate step for inclusion within the paper.
\begin{itemize}
    \item $\psi_1$ is defined as $\ket{z}\ket{(z-p_1) \times v_1}\ket{g(z, v_1, p_1)}$. Here, we evaluate the point $z$ against the line defined by $\{p_1 + \lambda v_1: \lambda \in \mathbb{R}\}$.
    \item For $\psi_2$, we have $\ket{z}\ket{(z-p_2) \times v_2}\ket{g(z, v_1, p_1) + g(z, v_2, p_2)}$. The comparison is now with the line $\{p_2 + \lambda v_2: \lambda \in \mathbb{R}\}$.
    \item In $\psi_3$, the state is $\ket{z}\ket{(z-p_n) \times v_n}\ket{\sum_{i=1}^{n} g(z, v_i, p_i)}$. The point $z$ is juxtaposed with the line $\{p_n + \lambda v_n: \lambda \in \mathbb{R}\}$.
    \item Finally, $\psi_4$ is represented as $\ket{z}\ket{0}\ket{\sum_{i=1}^{n} g(z, v_i, p_i)}$. The ancillary register values are reset to their initial state in the last step.
\end{itemize}

This sequence of states $\psi_1$ to $\psi_4$ articulates the computational procedure for determining the relative position of a point $z$ concerning a sequence of lines, each represented by a point $p_i$ and a direction vector $v_i$. The function $g$ computes the cross-product and contributes to a cumulative sum that ultimately discerns the point's inclusion within the polygon defined by these lines. The final step ensures the reversibility of the quantum operations by resetting the ancillary qubits.

\section{Results}\label{sec:results}

This section presents the empirical validation of our quantum algorithms designed to tackle spatial discretization in the presence of various geometric obstacles. Through a series of simulations and quantum circuit implementations, we demonstrate the effectiveness of our proposed methods across three distinct scenarios: rectangular in Fig. \eqref{fig:res_rectsngulars}, circular in Fig. \eqref{fig:res_circulars}, and convex polygonal obstacles in Fig. \eqref{fig:res_poligonals}. As a part of a result of our work, fig. \eqref{fig:performance_comparison} shows the quadratic improvement of the quantum algorithm over the classical approach.
The performance of each operator, integral to the function of the quantum oracles, is scrutinized to ascertain their contribution to the overall efficacy of the algorithm. As a result, tables \eqref{tab:quantum_operators_resources} and \eqref{tab:quantum_operators_desc} summarize the operators created as contributions and ingredients to respond to the problem of spatial discretization.

\subsection{Rectangular Obstacle Scenario}
\begin{figure}[]
\centering
\includegraphics[width=0.2\textwidth]{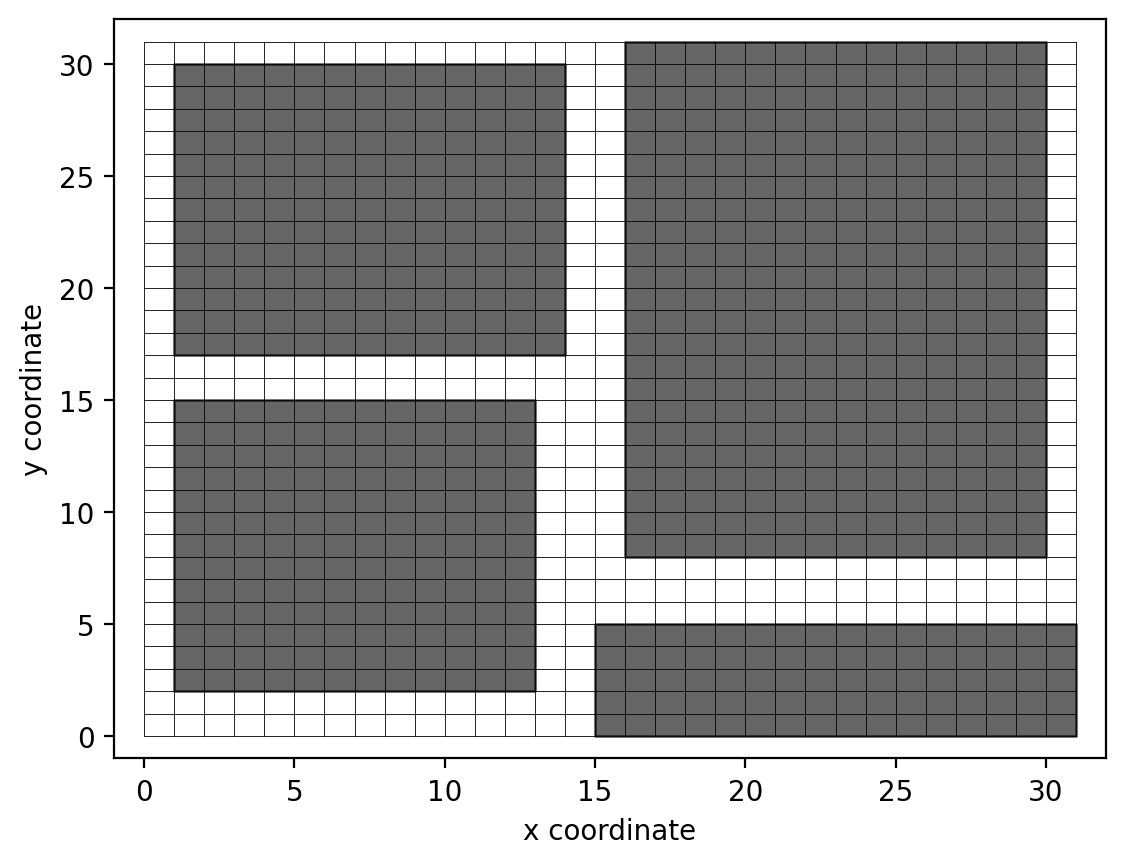}
\includegraphics[width=0.2\textwidth]{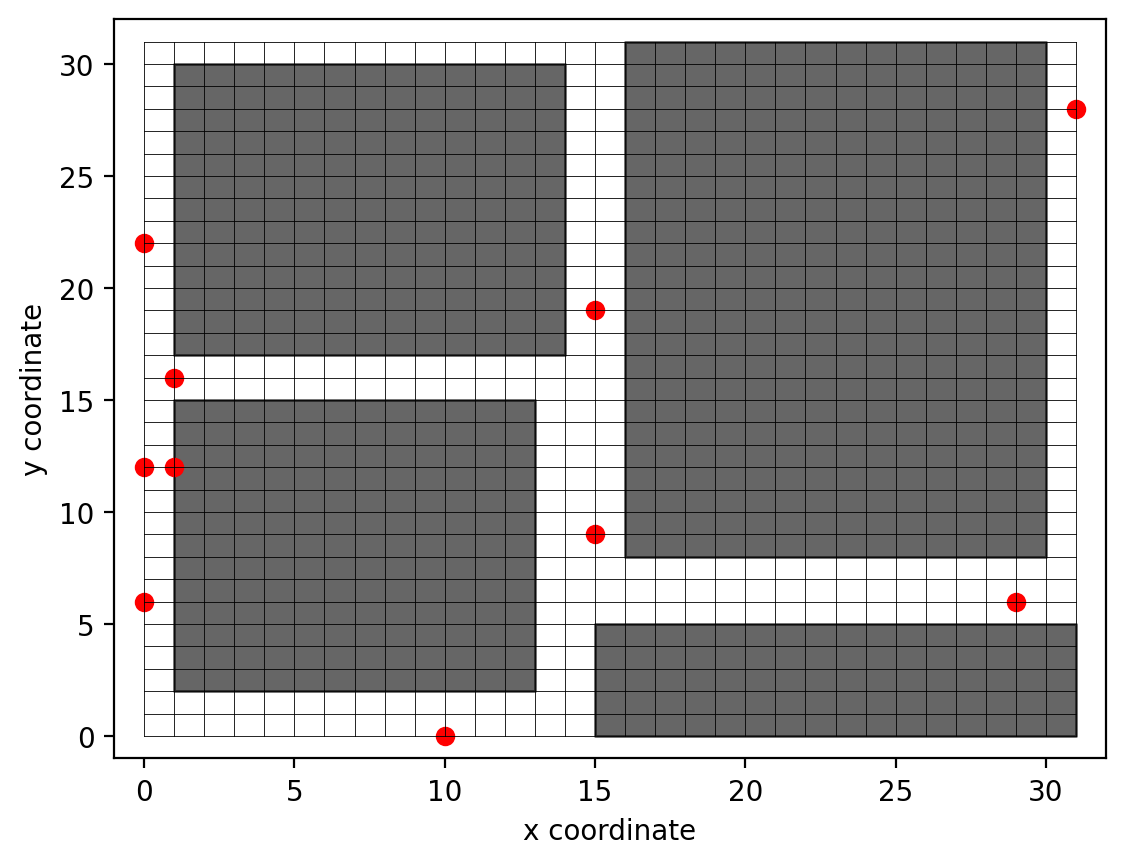}
\includegraphics[width=0.2\textwidth]{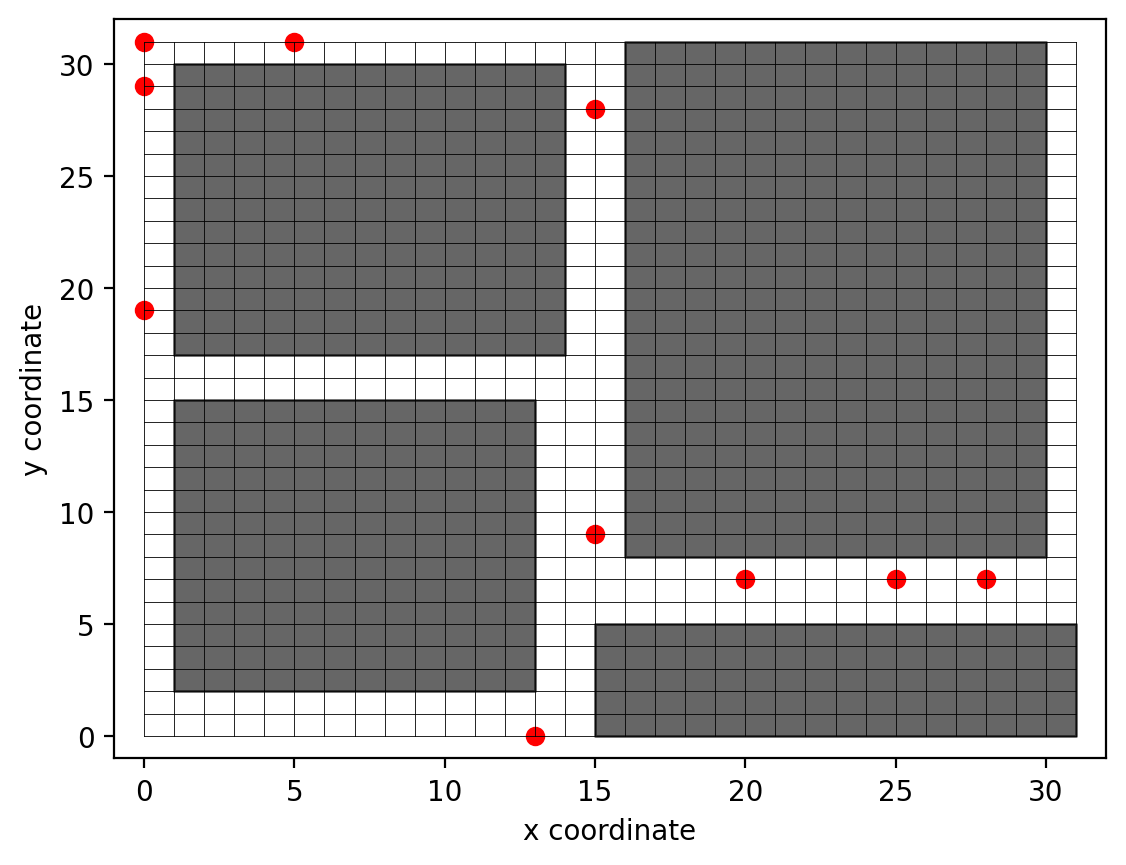}
\includegraphics[width=0.2\textwidth]{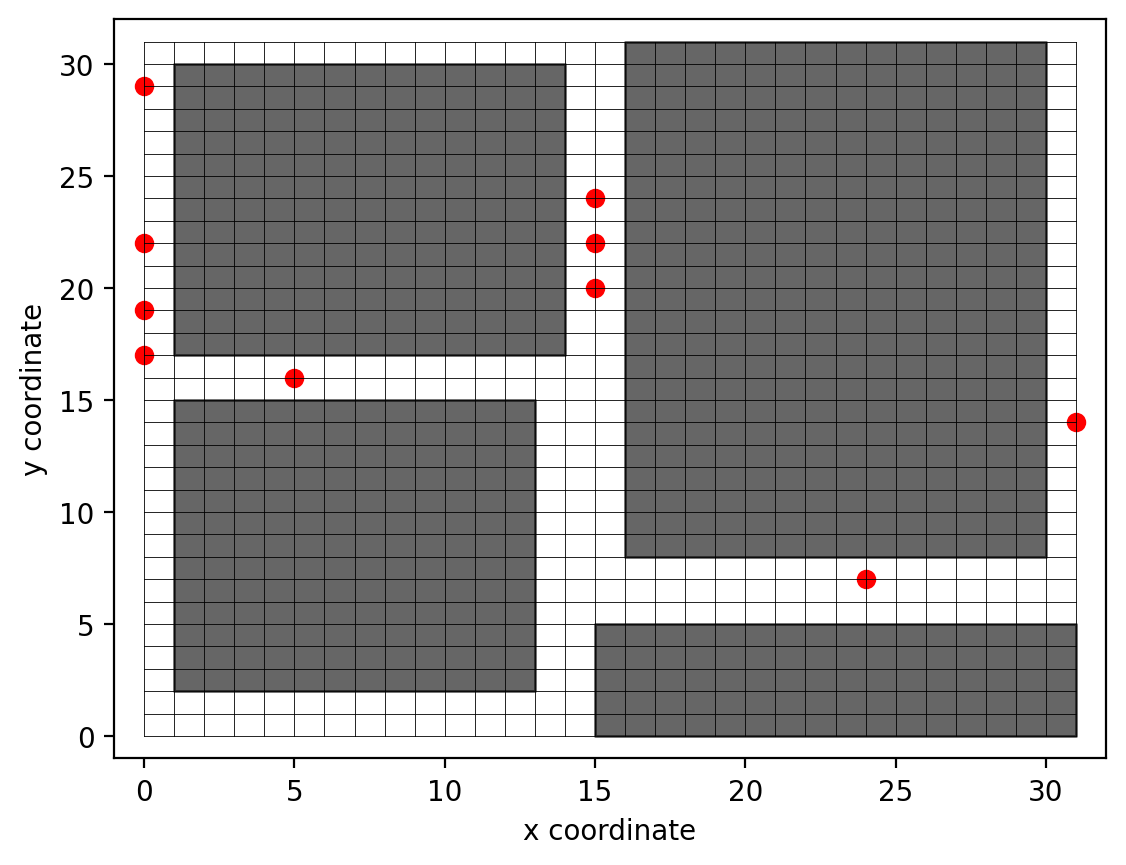}

\caption{This figure displays the various tests we have conducted. We selected ten red points that do not belong to the rectangles. We then applied Grover's algorithm and demonstrated that the points returned were within the feasible zone.}
\label{fig:res_rectsngulars}
\end{figure}

In the rectangular obstacle scenario, we utilized the \textit{Interval Check Operator} (ICO), \textit{Rectangle Inclusion Operator} (RIO), and \textit{Multiple Rectangles Inclusion Operator} (MRIO) within our quantum circuits. The \textit{ICO} accurately determined the inclusion of points within specified intervals, while the RIO successfully verified point inclusion within individual rectangles. The \textit{MRIO} extended this functionality to multiple rectangles, showcasing the scalability of our approach. The results figures can be analyzed in Fig. \eqref{fig:res_rectsngulars}.

\subsection{Circular Obstacle Scenario}
\begin{figure}[]
\centering
\includegraphics[width=0.2\textwidth]{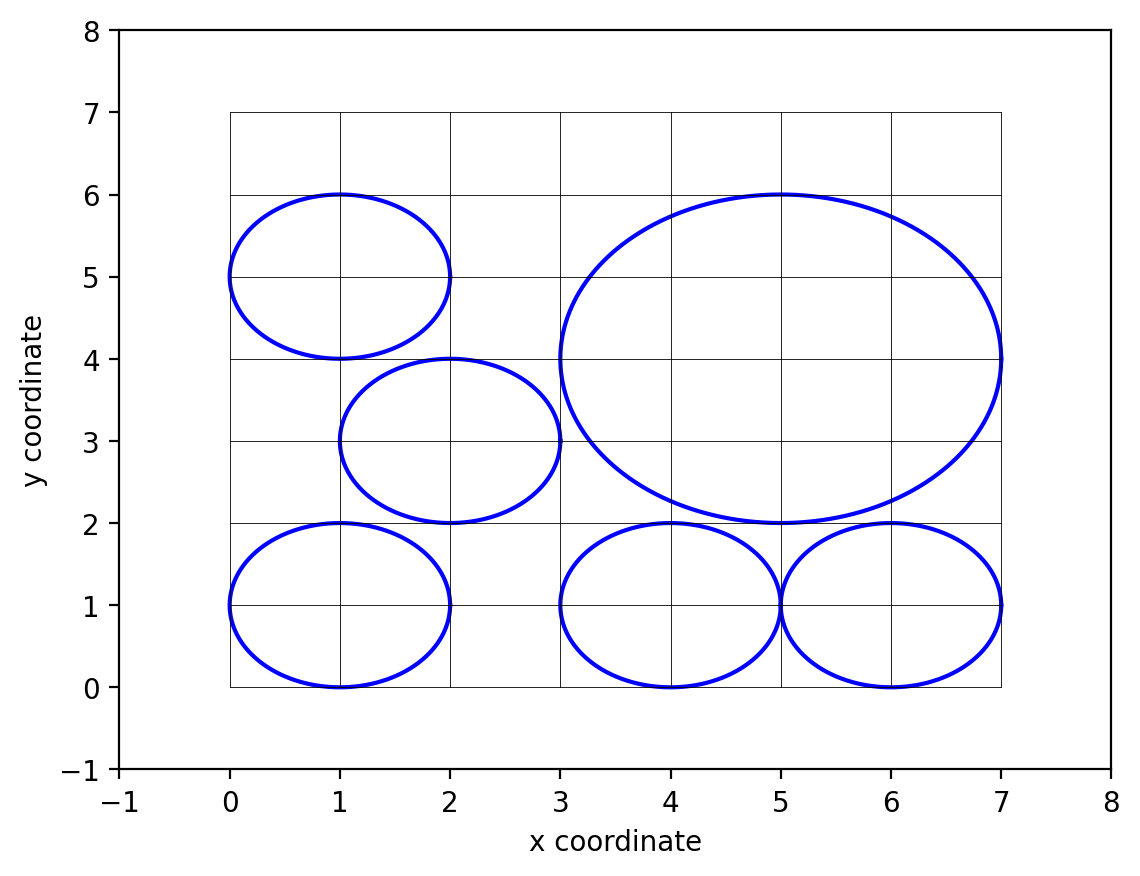}
\includegraphics[width=0.2\textwidth]{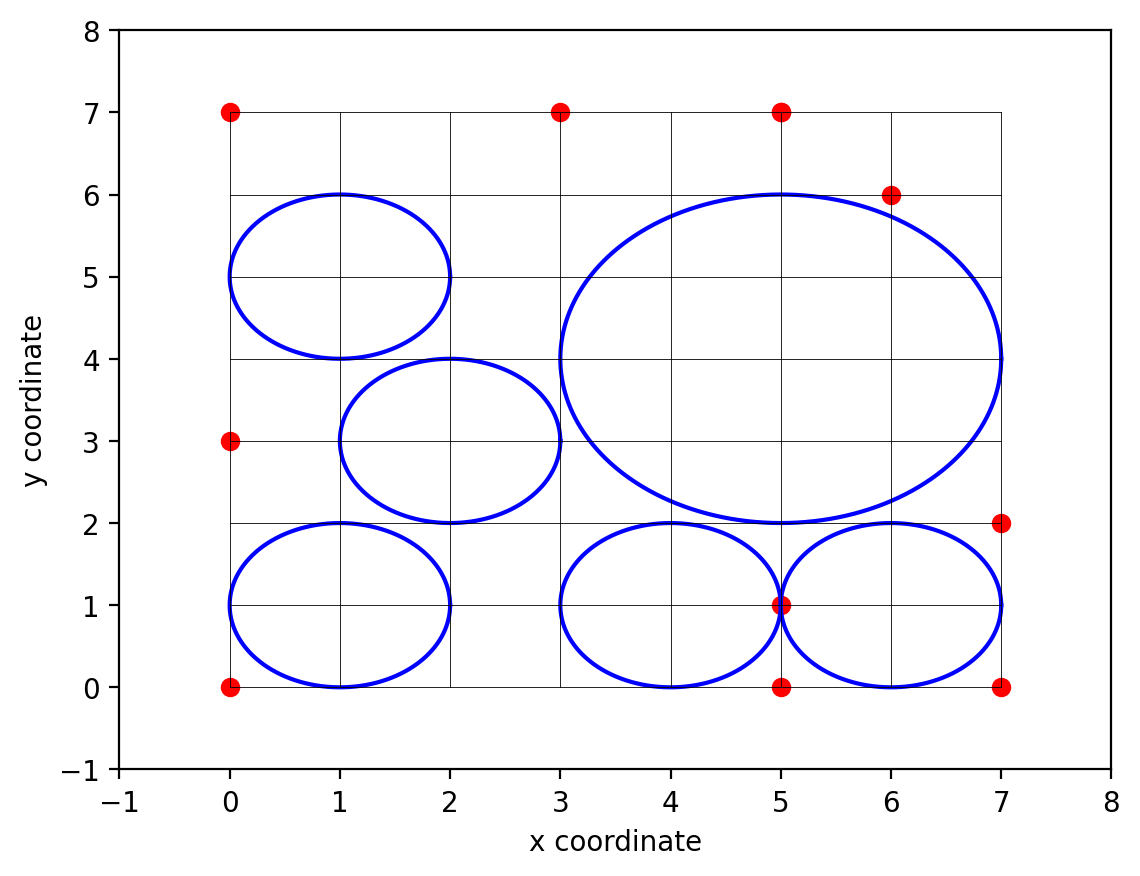}
\includegraphics[width=0.2\textwidth]{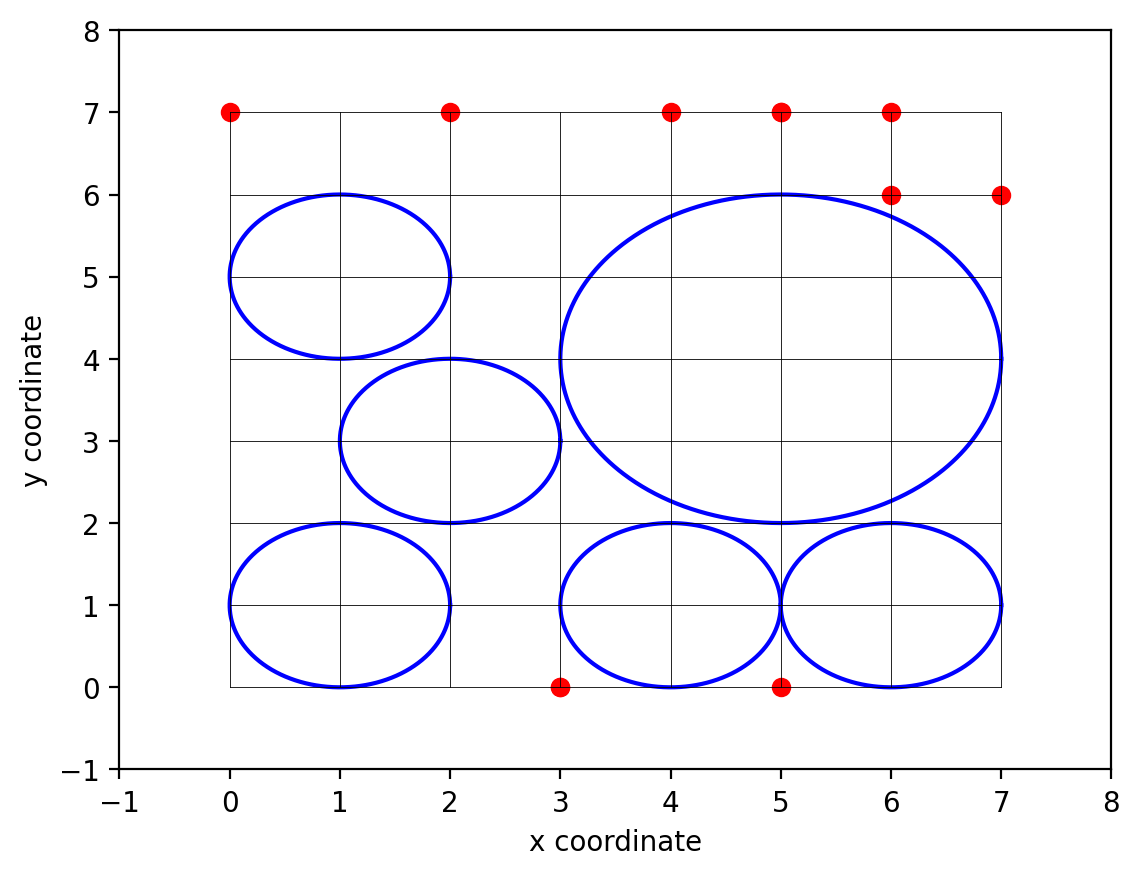}
\includegraphics[width=0.2\textwidth]{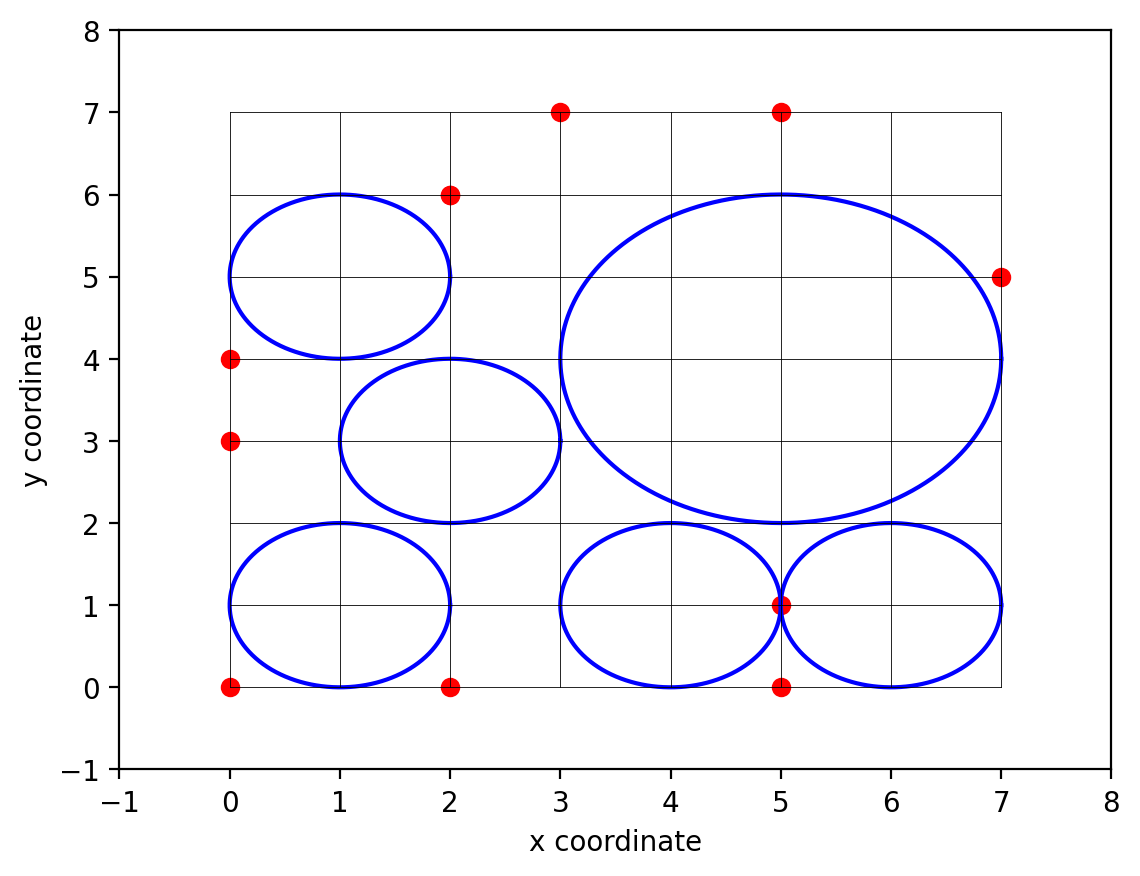}

\caption{This figure displays the various tests we have conducted. We selected nine red points that do not belong to the circular obstacle. We then applied Grover's algorithm and demonstrated that the points returned were within the feasible zone. }
\label{fig:res_circulars}
\end{figure}

For circular obstacles, the \textit{Quantum Threshold Comparator} (\(\hat{T}\)) and the \textit{Quantum Negation Operator} (\(\hat{N}\)) were pivotal. The \(\hat{T}\) efficiently encoded the binary variable \(\alpha\), indicating the relation between a point and the circle's center, while \(\hat{N}\) calculated the negation modulo \(2^n\), crucial for the distance computations. The operators collectively facilitated the construction of an oracle that accurately discerned point exclusion from circular areas. The circular obstacle scenario's results figures can be analyzed in Fig. \eqref{fig:res_circulars}.

\subsection{Convex Polygonal Obstacle Scenario}
\begin{figure}[]
\centering
\includegraphics[width=0.2\textwidth]{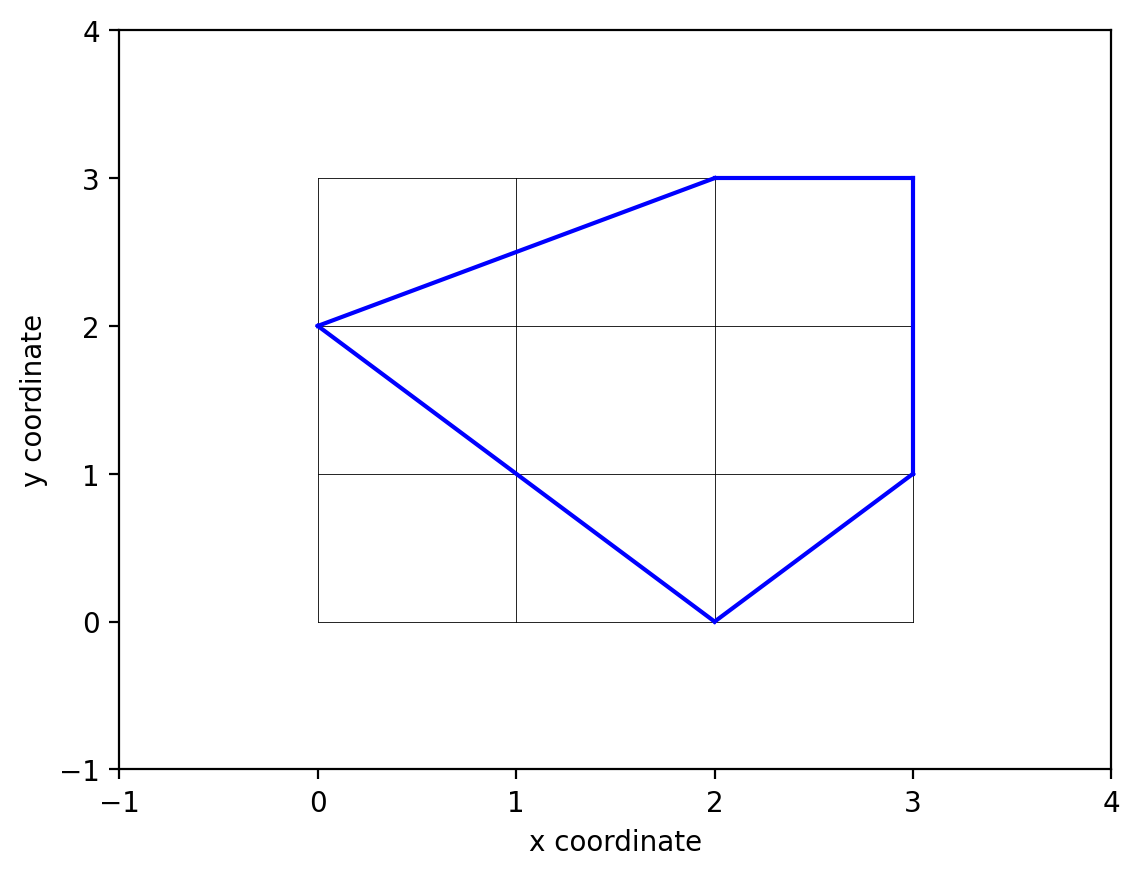}
\includegraphics[width=0.2\textwidth]{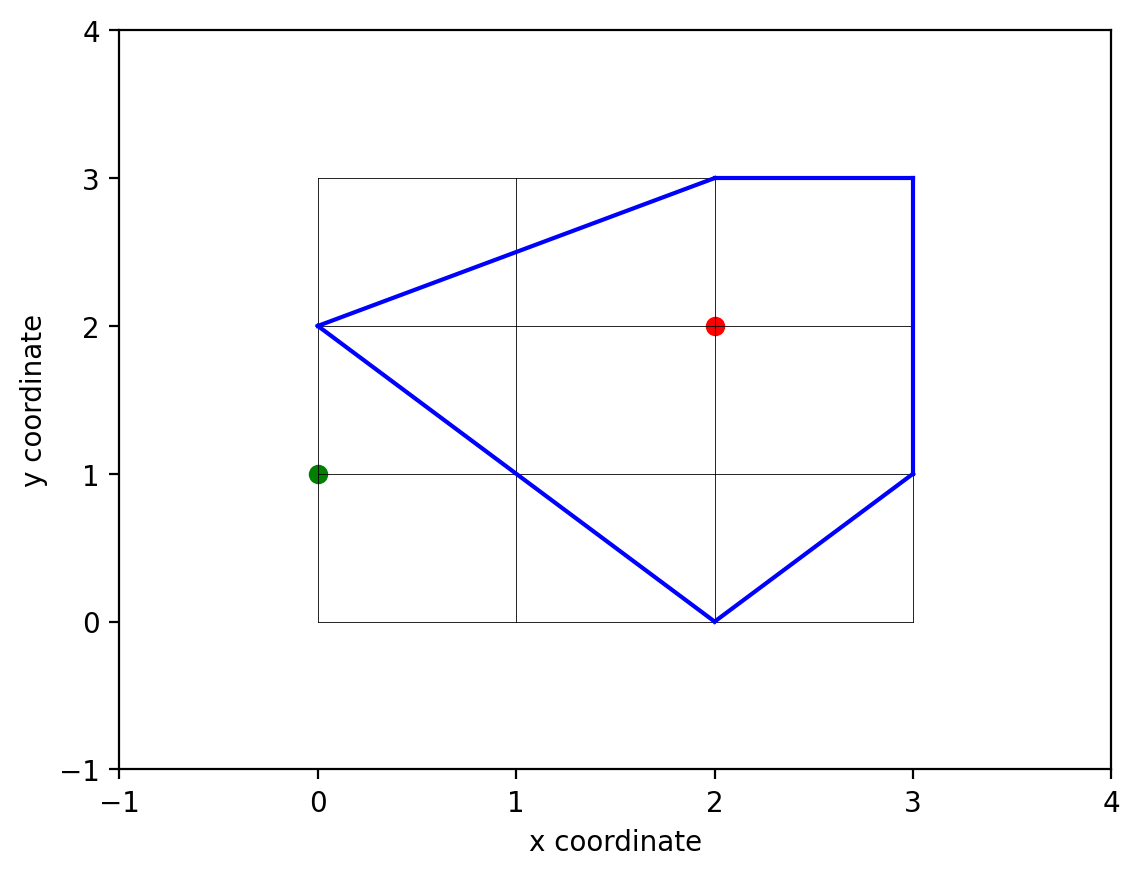}
\includegraphics[width=0.2\textwidth]{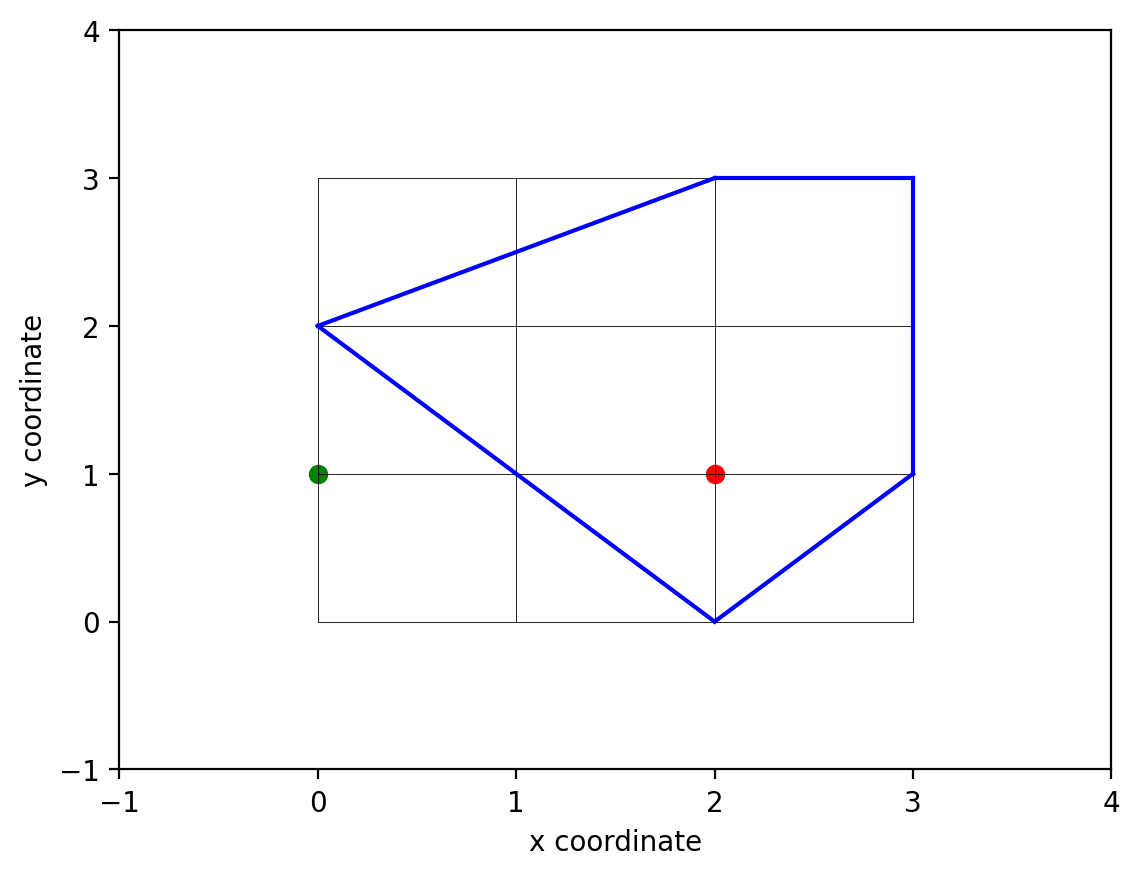}
\includegraphics[width=0.2\textwidth]{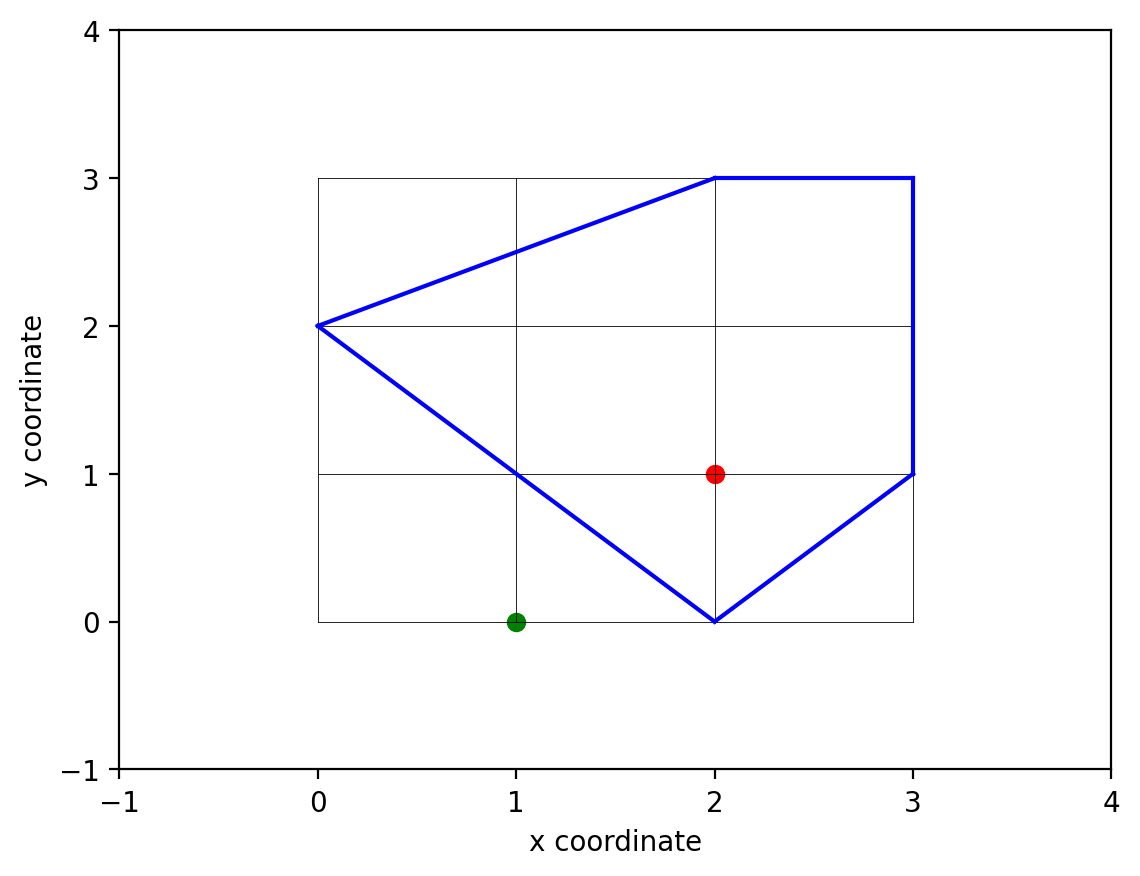}

\caption{This figure displays the various tests we have conducted. We selected 1 red point that does belong to the polygon and tested that the green point was within the feasible zone. }
\label{fig:res_poligonal}
\end{figure}

\begin{figure}[]
\centering
\includegraphics[width=0.2\textwidth]{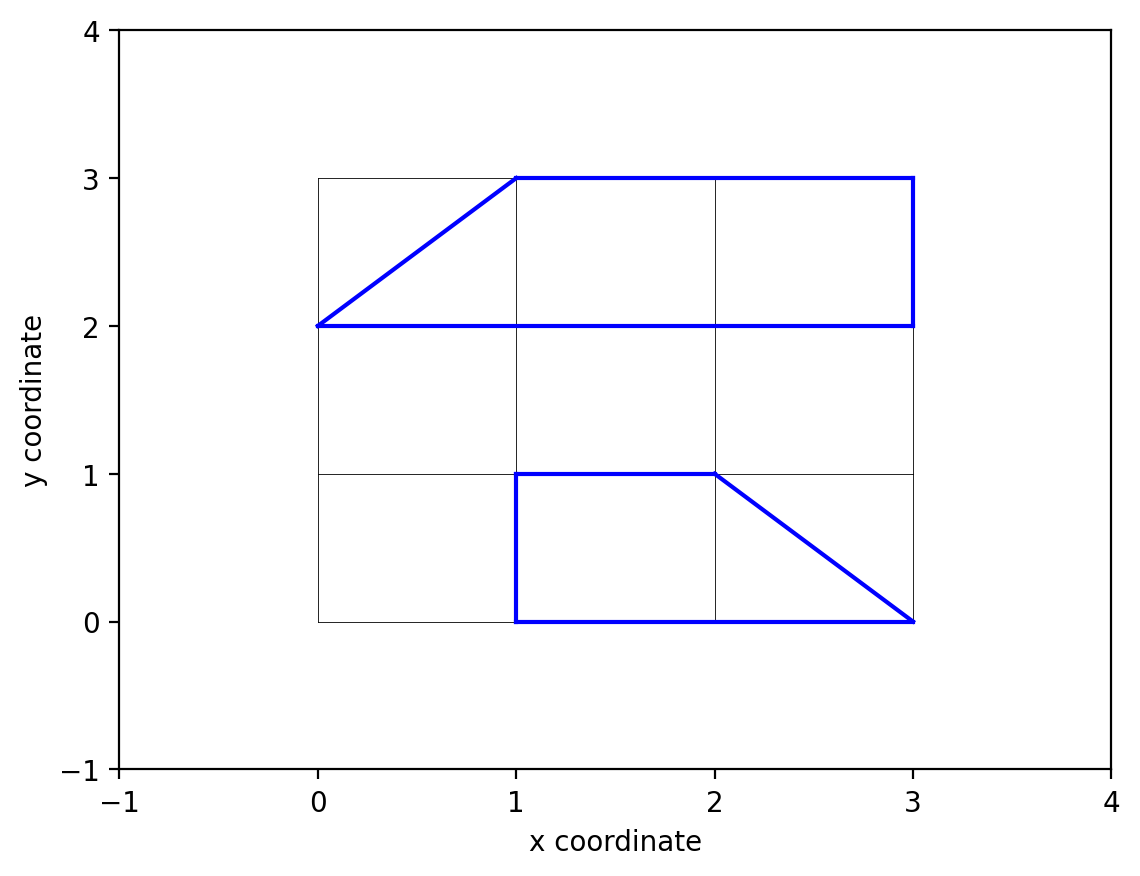}
\includegraphics[width=0.2\textwidth]{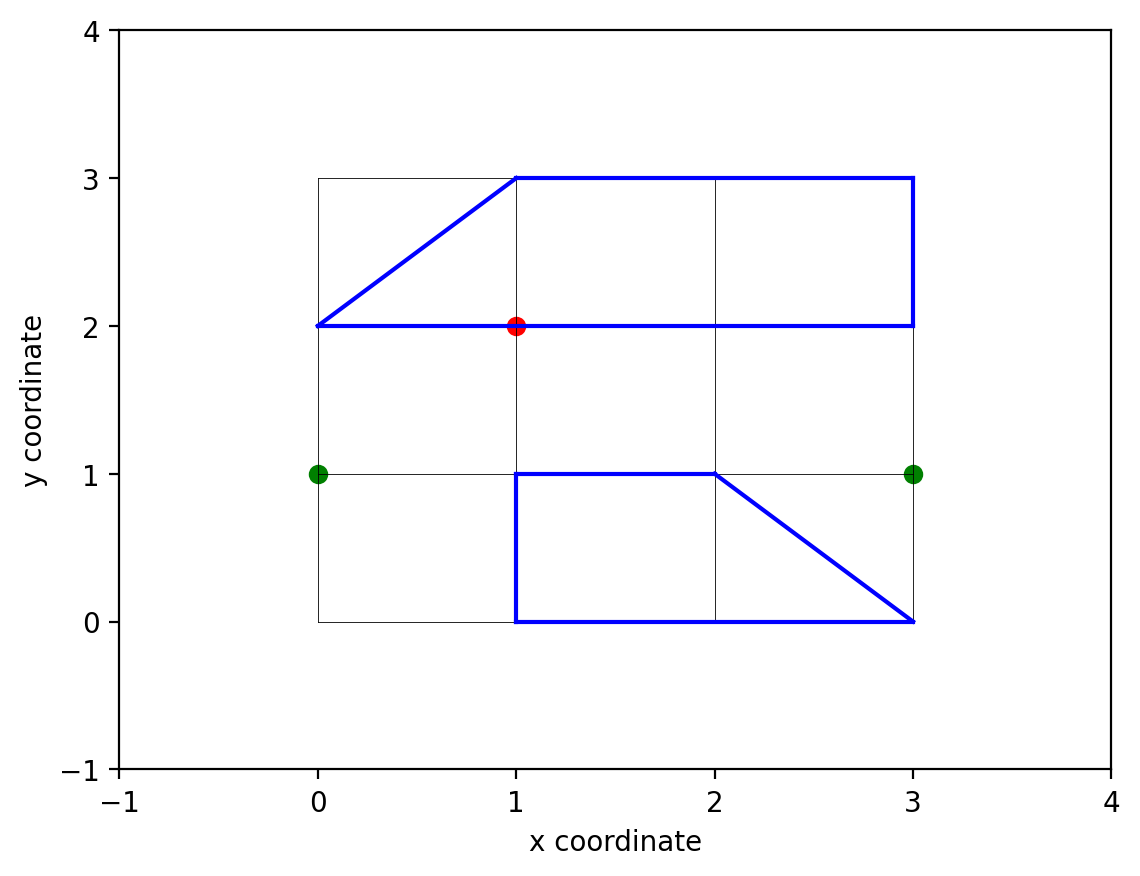}
\includegraphics[width=0.2\textwidth]{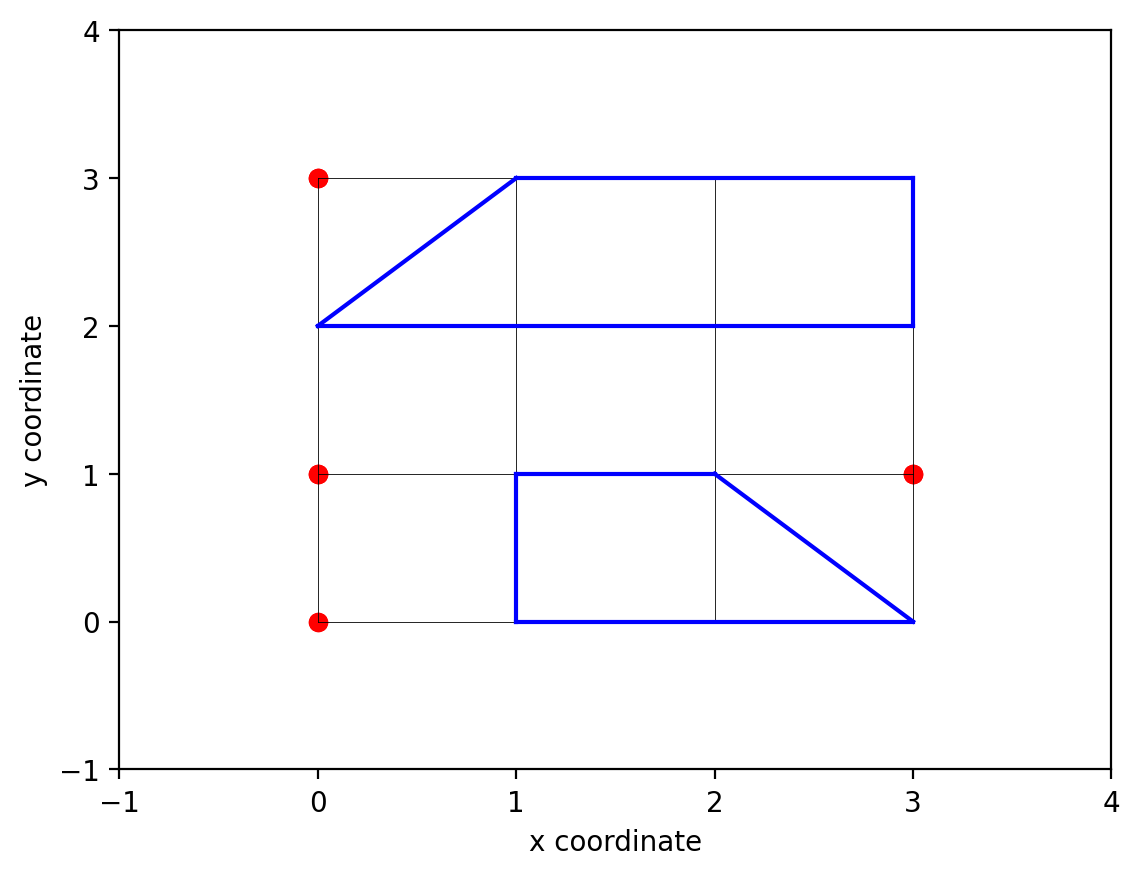}
\includegraphics[width=0.2\textwidth]{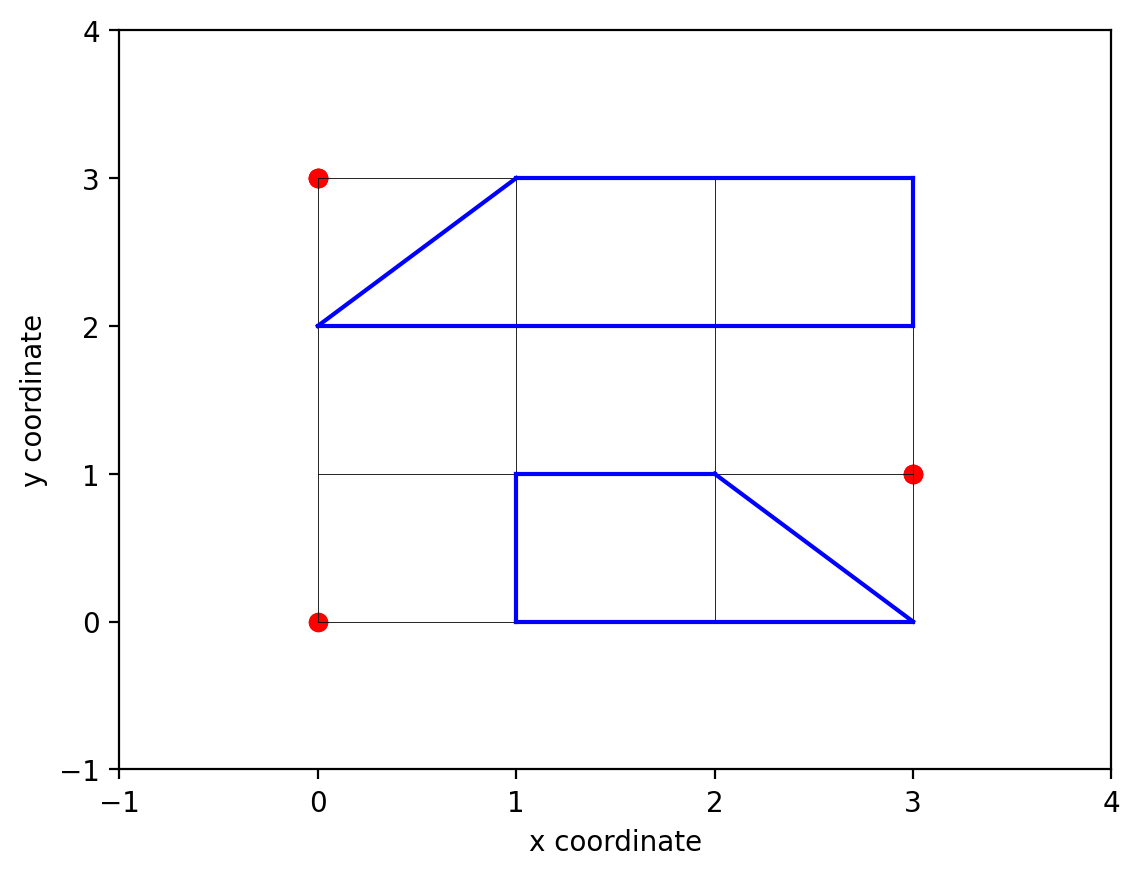}
\caption{This figure displays the various tests we have conducted. We selected 3 red points that do not belong to the polygonal obstacle. We then applied Grover's algorithm and demonstrated that the points returned were within the feasible zone. }
\label{fig:res_poligonals}
\end{figure}

In dealing with convex polygonal obstacles, the quantum circuit effectively computed vector cross-products using the \textit{Quantum Adder Operator} (Add(k)) and the \textit{Quantum Multiplication Operator}(Mult(k)). These operators allowed the circuit to evaluate whether a point resided within the polygon by checking the sign consistency of cross-products, which is a quantum computational reflection of the left-hand rule.
Fig. \eqref{fig:res_poligonal} and \eqref{fig:res_poligonals} show the convex polygonal obstacle scenario's results figures.

The diverse range of operators implemented and tested herein lays the foundation for more complex geometric processing tasks within the quantum domain. Our findings not only validate the utility of these operators but also serve as a testament to the potential of quantum algorithms to empower computational geometry.

\begin{table*}[ht]
\centering
\caption{Quantum operators implemented in the algorithm with their respective resource requirements. The number of qubits required, the count of auxiliary qubits, and the depth of each operator's circuit are crucial factors for assessing the algorithm's complexity and feasibility on quantum hardware. The number of rectangular obstacles is the variable $m$ in the $MRIO$ operator.}
\label{tab:quantum_operators}
\begin{tabular}{ccccc}
\toprule
\hline
\textbf{Operator} & \textbf{Input Qubits} & \textbf{Output Qubits } & \textbf{Auxiliary Qubits} & \textbf{Depth} \\ 
\hline
\midrule
$ICO(a_1, a_2)$ & \(n\) & 1 & 3 & \(O(n^2)\) \\ 
$RIO(R)$ & \(n\) & 1 & 5 & \(O(n^2)\) \\ 
$MRIO(\{R_i\}_{i=1}^m)$ & \(n\) & \(\log_2(m)\) & 6 & \(O(n^2)\) \\ 
$\hat{T}$ & \(n\) & 1 & 1 & \(O(n^2)\) \\ 
$\hat{N}$ & \(n+1\) & 0 & 0 & \(O(n^2)\) \\ 
$Mult$(k) & \(n\) & \(2n\) & 0 & \(O(n^2)\) \\ 
$\textit{Add\_{in}}(k)$ & \(n\) & 0 & 0 & \(O(n^2)\) \\ 
$Add\_{in}$ & \(n\) & 0 & 0 & \(O(n^2)\) \\ 
$AbsDiff(k)$ & \(n+1\) & \(n+1\) & 2 & \(O(n^2)\) \\ 
$AddSqr_{in}$ & \(n\) & \(2n\) & 0 & \(O(n^2)\) \\ 
$ED_{in}(c_1,c_2)$ & \(2n\) & \(2n\) & $n+1$ & \(O(n^2)\) \\ 
$VCP(v_1,v_2)$ & \(2n\) & \(2n\) & 1 & \(O(n^2)\) \\ \hline
\end{tabular}
\label{tab:quantum_operators_resources}
\end{table*}

\section{Discussion}\label{sec:discussion}

\begin{figure}[]
\centering
\includegraphics[width=0.5\textwidth]{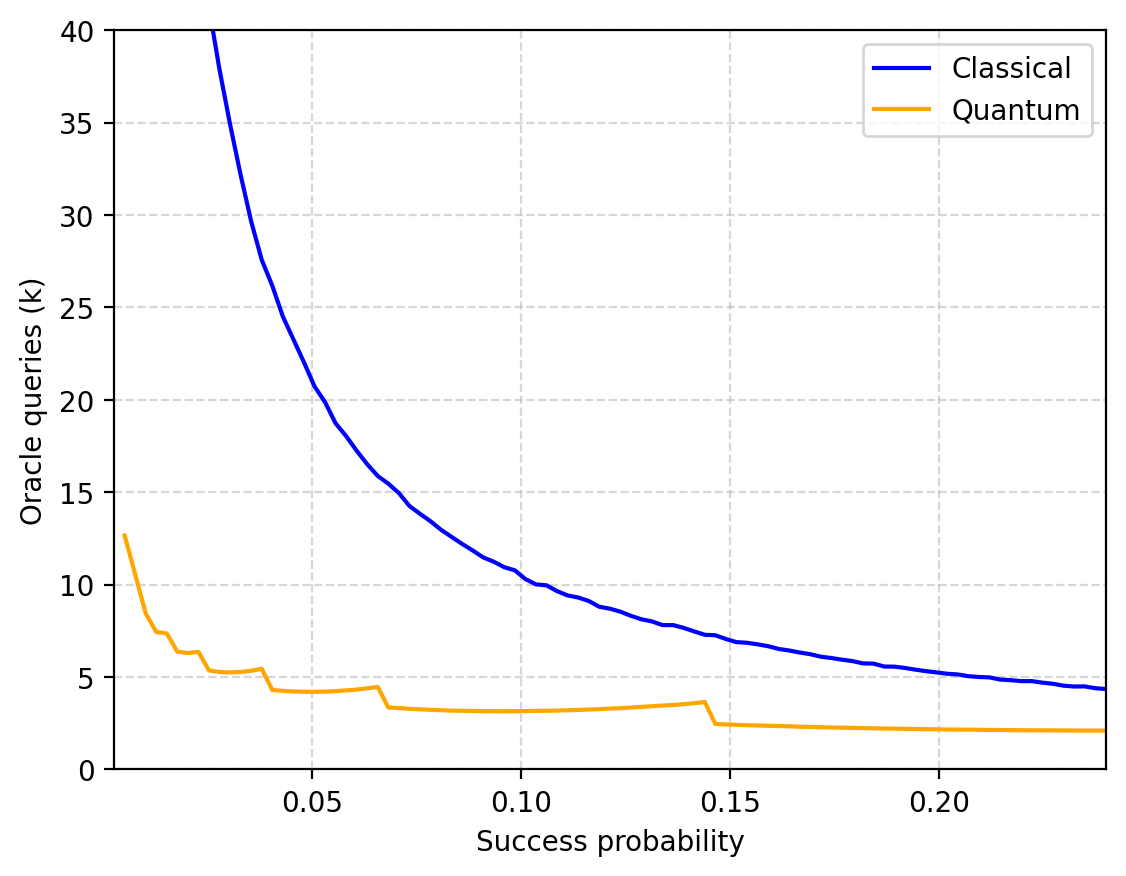}
\caption{Comparative performance graph showing the quadratic improvement of the quantum algorithm over the classical approach in the task of Spatial discretization. The x-axis represents the input size, while the y-axis indicates the number of Oracle applications required. The quantum algorithm's curve illustrates a markedly lower growth rate, confirming its superior efficiency.}
\label{fig:performance_comparison}
\end{figure}

This section elucidates the significant advancements made in the field of quantum algorithmic geometry through the development of our quantum algorithm. A critical aspect of our research is demonstrating the quadratic improvement of our algorithm's performance over classical methods. We have substantiated this improvement through a rigorous analytical framework and simulations, depicted in a comparative graph (see Figure \eqref{fig:performance_comparison}). We also provide a demonstration (\eqref{sec:appendix}), showcasing the quadratic improvement that is reflected in the comparative Fig. \eqref{fig:performance_comparison}

Implementing our quantum algorithm involved developing a series of quantum operators, each integral to the spatial discretization process. These operators were ingeniously crafted to align with the capabilities and limitations of current quantum computing technology, ensuring both theoretical robustness and practical applicability. Table \ref{tab:quantum_operators} provides a comprehensive summary of these operators, detailing the number of qubits required, their specific roles within our algorithm, and their computational depth. Notably, the computational depth, predominantly $O(n^2)$ due to applying the \textit{Quantum Fourier Transform} (QFT), could be optimized for the actual quantum era. Our scope is the \textit{Intermediate-Scale Quantum} (ISQ) systems \cite{FromNISQtoISQs}. Our approach thus stands as a testament to the adaptability and scalability of quantum algorithms in tackling complex geometric problems.

\begin{table*}[ht]
\centering
\begin{tabular}{ccp{0.6\textwidth}}
\hline
\textbf{Operator} & \textbf{Symbol} & \textbf{Description} \\
\hline
Interval Check Operator & \( ICO (a_1,a_2)\) & Determines if a given number falls within a specified interval, crucial for boundary verification. \\
\hline
Rectangle Inclusion Operator & \( RIO (R) \) & Verifies the inclusion of a point within the boundaries of a rectangle, fundamental in obstacle avoidance. \\
\hline
Multiple Rectangles Inclusion Operator & \( MRIO(\{R_i\}_{i=1}^m) \) & Extends the functionality of \( RIO \) to assess inclusion within multiple rectangular obstacles simultaneously. \\
\hline
Quantum Threshold Comparator & \( \hat{T} \) & Evaluates whether a given value exceeds a specified threshold, used in distance calculations. \\
\hline
Quantum Negation Operator & \( \hat{N} \) & Performs negation operations on quantum states, essential for computing absolute differences. \\
\hline
Inplace Quantum Adder & \( Add_{\text{in}}(k) \) & Adds a constant \( k \) to a quantum state, crucial for cumulative calculations. \\
\hline
General Quantum Adder & \( Add_{\text{in}} \) & Facilitates general addition operations between quantum states. \\
\hline
Absolute Difference Operator & \( AbsDiff (k) \) & Computes the absolute difference between quantum states, significant in distance evaluations. \\
\hline
Quantum Square Adder & \( AddSqr_{\text{in}} \) & Executes the square addition operation, pivotal in distance and area calculations. \\
\hline
Quantum Euclidean distance & \( ED_{\text{in}}(c_1,c_2) \) & Executes the Euclidean distance. \\
\hline
Quantum Vector Cross Product & \( VCP \) & Executes the vector cross product between two vectors. \\
\hline
\end{tabular}
\caption{Summary of quantum operators developed for the quantum algorithm. Each operator is integral to the various stages of spatial discretization in the presence of obstacles.}
\label{tab:quantum_operators_desc}
\end{table*}

With the improvement shown with the proposed algorithm, we pretend to push the boundaries of what is achievable with quantum computing; the implications of this research extend far beyond the immediate results, promising a new horizon of possibilities in various applications requiring spatial analysis and path planning.

\section{conclusion}\label{sec:conclusions}
Our work introduces an innovative quantum algorithm designed to address the generalized problem of spatial discretization, particularly in environments peppered with obstacles of varying geometries, such as rectangles, circles, and polygons. At the core of our algorithm lies an array of meticulously crafted quantum operators, each tailored to navigate the complexities of such spatial arrangements. Our approach not only achieves a quadratic enhancement in efficiency compared to classical methods but also signifies a remarkable advancement in computational geometry and quantum computing. The pronounced superiority of our quantum algorithm, in terms of performance, is compellingly illustrated in Figure \eqref{fig:performance_comparison}, underscoring its potential to revolutionize spatial analysis in complex settings.

The quantum operators developed, such as the \textit{Interval Check Operator} (ICO), \textit{Rectangle Inclusion Operator} (RIO), and \textit{Quantum Threshold Comparator} (\(\hat{T}\)), among others, are instrumental in processing and analyzing spatial data with unparalleled efficiency. These operators showcase the potential of quantum computing to transform tasks that are traditionally challenging due to computational intensity and precision requirements.

Our empirical results, obtained through simulations in the PennyLane quantum software library, validate the practical effectiveness of our algorithm. The comparative analysis against classical approaches highlights the quantum advantage regarding speed and efficiency and demonstrates the algorithm’s viability within the current quantum computing landscape \eqref{fig:performance_comparison}.

As we stand at the cusp of the quantum era, this research contributes significantly to the emerging field of quantum computational geometry. It paves the way for future explorations and applications of quantum algorithms in various domains requiring spatial discretization. The implications of this study extend beyond theoretical advancements; they offer insights into potential real-world applications, from urban planning and autonomous navigation to complex simulations in medicine and machine learning.

Our work underscores the transformative potential of quantum algorithms in addressing complex geometric problems. It opens new horizons for quantum-enhanced solutions in computational tasks, setting a benchmark for future research in quantum computational geometry and related fields.

\section*{Code}\label{sec:code}
The Python code implementing the operators and reproducing the figures discussed herein is available in the accompanying GitHub repository. The code provides the foundational algorithms enabling in-depth exploration and adaptation of the operators. The code can be accessed at \url{https://github.com/pifparfait/Quantum-improvement-in-Spatial-Discretization}.

\section*{Acknowledgements} The authors thank Guillermo Alonso de Linaje for the discussions and consideration during the experiments.

\section*{Compliance with Ethics Guidelines}
Funding: This research received no external funding. 
Institutional review: This article contains no studies with human or animal subjects.
Informed consent: Informed consent was obtained from all participants in the study.
Data availability: Data sharing is not applicable. No new data were created or analyzed in this study. Data sharing does not apply to this article.

\appendix
\section{Comparative Analysis of Classical and Quantum Oracle Efficacy}\label{sec:appendix}
 We compare the application frequency of the function $f$, which determines point inclusion within the feasible zone for both classical and quantum cases. We aim to acquire $M = 1000$ points within the feasible zone, where $p$ represents the probability of randomly selecting a point within this zone. Given the total area $A_T$ and the feasible area $A_F$, the probability $p$ is given by $p = \frac{A_F}{A_T}$. Here, we calculate each case's expected number of function applications $f$.

\subsection{Classical Case}
In the classical scenario, we want to sample $M$ points within the feasible zone, with $p$ being the probability of a point falling. The expected number of Oracle applications is $\frac{M}{p}$.

If $X_i$ is a Bernoulli random variable taking values in $\{0,1\}$ with probabilities $\{1-p, p\}$, the distribution modeling the required number of trials $N$ to achieve $M$ successes is as follows. The probability that exactly $N$ trials are necessary is given by:

\begin{equation}
    P(Y=N) = \binom{N-1}{M-1}p^{M}(1-p)^{N-M},
    \label{eq:probability_Y}
\end{equation}
which we reference as the probability distribution of $Y$ (Equation \ref{eq:probability_Y}).

Calculating the expected value, we have:

\begin{align}
    E[Y] &= \sum_{N = M}^\infty N P(Y = N) \nonumber \\
    &= \sum_{N = M}^\infty N \binom{N-1}{M-1}p^{M}(1-p)^{N-M} \nonumber \\
    &= \frac{Mp}{(1-p)^{M+1}}\sum_{N = M}^\infty \binom{N}{M}(1-p)^{N}, \label{eq:expected_Y_calculation}
\end{align}
where we apply the binomial theorem to arrive at the expected value (Equation \ref{eq:expected_Y_calculation}).

\begin{equation}
    \sum_{N=M}^\infty \binom{N}{M}(1-p)^{N} = \frac{(1-p)^M}{p^{M+1}}.
    \label{eq:binomial_theorem_application}
\end{equation}

Combining \eqref{eq:expected_Y_calculation} to \eqref{eq:binomial_theorem_application}, we arrive at:

\begin{equation}
    E[Y] = \frac{M}{p}.
    \label{eq:classical_num_queries}
\end{equation}
This is the expected number of queries in the classical case (Equation \ref{eq:classical_num_queries}).

\subsection{Quantum Case}
In the quantum scenario, detailed in Equation \ref{eq:quantum_num_queries}, we seek to sample $M$ points within the feasible zone with probability $p$. Let $CI(x)$ denote the closest integer to a real number $x$, and let $\phi := \arcsin(\sqrt{2p(1-p)})$. According to Grover's algorithm theory \cite{grover1996fast}, after applying the oracle $R$ times with $R := CI\left(\frac{\arccos(\sqrt{p})}{\phi}\right)$, the probability of finding a point in the correct zone is close to $1$. Hence, the number of times we need to apply the oracle $f$ is $(R+1)*M$, with $R \leq \sqrt{\frac{1}{p}}$. This yields:

\begin{equation}
    (R+1)M \thickapprox \frac{M}{\sqrt{p}} < \frac{M}{p},
    \label{eq:quantum_num_queries}
\end{equation}
which is a quadratic improvement over the $\frac{M}{p}$ \eqref{eq:classical_num_queries} applications required by the classical algorithm (Equation \ref{eq:quantum_num_queries}).

\newpage

\bibliography{main} 
\end{document}